\documentclass[aps,prb,twocolumn,showkeys,superscriptaddress,hidelinks]{revtex4}
\usepackage{graphics}
\usepackage{graphicx}
\usepackage{array}
\usepackage{dcolumn}
\usepackage{bm}
\usepackage{float}
\usepackage{enumitem}
\usepackage[colorinlistoftodos, textwidth=4cm, shadow]{todonotes}
\usepackage{amssymb}
\usepackage{hyperref}

\newcommand{\kB}{k$_{\rm B}$}
\newcommand{\MST}{Mn$_3$Si$_2$Te$_6$\,}
\newcommand{\TC}{$T_{\rm C}$\,}
\newcommand{\TN}{$T_{\rm N}$\,}

\begin{document}


\title{Magnetic order and interactions in ferrimagnetic Mn$_3$Si$_2$Te$_6$}

\author{Andrew F. May}
\email{mayaf@ornl.gov}
\affiliation{Materials Science and Technology Division, Oak Ridge National Laboratory, Oak Ridge, TN 37831}
\author{Yaohua Liu}
\affiliation{Quantum Condensed Matter Division, Oak Ridge National Laboratory, Oak Ridge, TN 37831}
\author{Stuart Calder}
\affiliation{Quantum Condensed Matter Division, Oak Ridge National Laboratory, Oak Ridge, TN 37831}
\author{David S. Parker}
\affiliation{Materials Science and Technology Division, Oak Ridge National Laboratory, Oak Ridge, TN 37831}
\author{Tribhuwan Pandey}
\affiliation{Materials Science and Technology Division, Oak Ridge National Laboratory, Oak Ridge, TN 37831}
\author{Ercan Cakmak}
\affiliation{Materials Science and Technology Division, Oak Ridge National Laboratory, Oak Ridge, TN 37831}
\author{Huibo Cao}
\affiliation{Quantum Condensed Matter Division, Oak Ridge National Laboratory, Oak Ridge, TN 37831}
\author{Jiaqiang Yan}
\affiliation{Materials Science and Technology Division, Oak Ridge National Laboratory, Oak Ridge, TN 37831}
\author{Michael A. McGuire}
\affiliation{Materials Science and Technology Division, Oak Ridge National Laboratory, Oak Ridge, TN 37831}

\date{\today}

\begin{abstract}
The magnetism in \MST has been investigated using thermodynamic measurements, first principles calculations, neutron diffraction and diffuse neutron scattering on single crystals.  These data confirm that \MST is a ferrimagnet below \TC$\approx$78\,K.  The magnetism is anisotropic, with magnetization and neutron diffraction demonstrating that the moments lie within the basal plane of the trigonal structure.  The saturation magnetization of $\approx$1.6$\mu_B$/Mn at 5\,K originates from the different multiplicities of the two antiferromagnetically-aligned Mn sites.  First principles calculations reveal antiferromagnetic exchange for the three nearest Mn-Mn pairs, which leads to a competition between the ferrimagnetic ground state and three other magnetic configurations.  The ferrimagnetic state results from the energy  associated with the third-nearest neighbor interaction, and thus long-range interactions are essential for the observed behavior.  Diffuse magnetic scattering is observed around the 002 Bragg reflection at 120\,K, which indicates the presence of strong spin correlations well above \TC.  These are promoted by the competing ground states that result in a relative suppression of \TC, and may be associated with a small ferromagnetic component that produces anisotropic magnetism below $\approx$330\,K.
\end{abstract}

\maketitle

\section{Introduction}

\href{https://doi.org/10.1103/PhysRevB.95.174440}{https://doi.org/10.1103/PhysRevB.95.174440}

Understanding and manipulating magnetic anisotropy is important in both basic and applied physics research.  For instance, anisotropic magnetism is essential for the development of permanent magnets, and it appears to be a fundamental component in many systems that display unconventional superconductivity.  Recently, research on materials that are crystallographically layered (held together by van der Waals bonds) has gained prominence in condensed matter physics.  This is largely driven by the now-realized prospect of building van der Waals heterostructures from materials with complementary properties.\cite{Geim2013} Two-dimensional and quasi-2D materials are of fundamental interest from a bulk perspective as well, in part because they exist in the limit of very anisotropic interactions (strong in-plane interactions and weak cross-plane interactions).  For instance, the compounds CrSiTe$_3$, MnPS$_3$, and CrI$_3$ have layers connected by van der Waals bonds, and demonstrate bulk magnetic ordering with anisotropic interactions manifesting themselves in anisotropic properties, suppressed 3D ordering temperatures, two-dimensional order, and persistent short-range correlations above \TC.\cite{Carteaux1995,Casto2015,William2015,Joy1992,Wildes1994,Okuda1986,Dillon1965,McGuire2015}  
In this work, we probe the behavior of \MST, which can be considered as a three-dimensional analogue of CrSiTe$_3$.

\begin{figure}[h!]%
\includegraphics[width=\columnwidth]{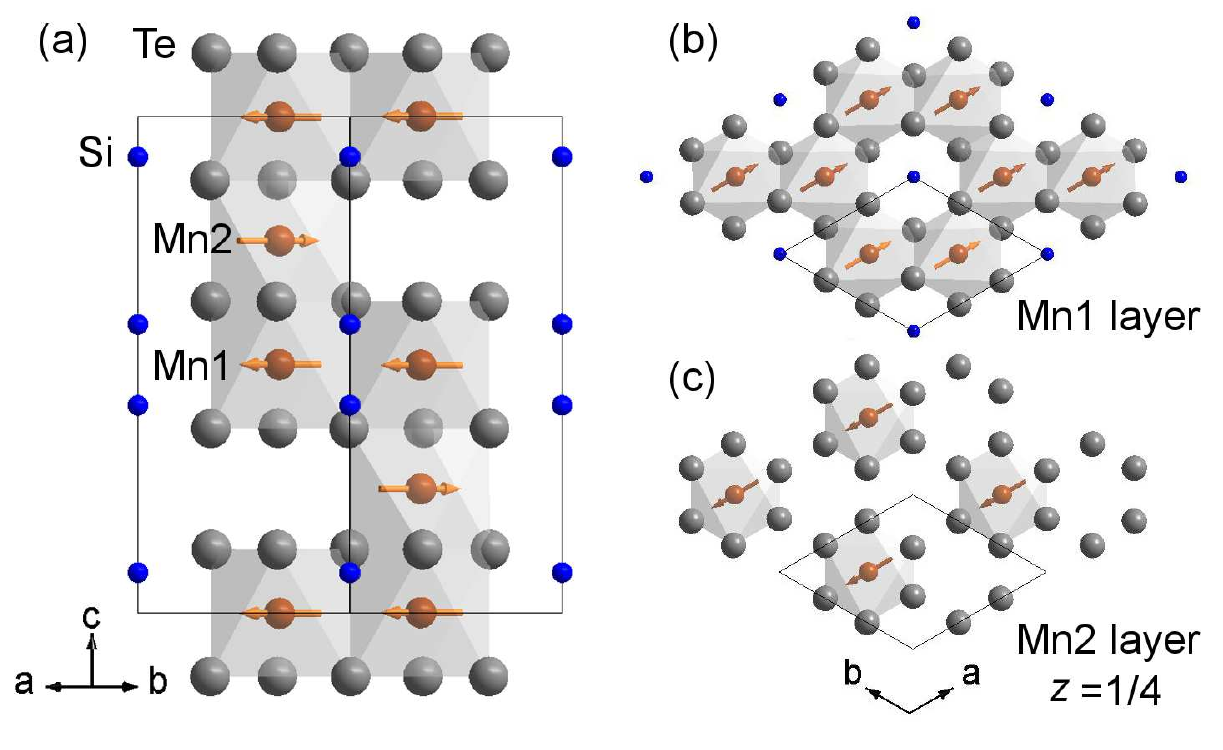}%
\caption{(a) Crystal structure of \MST viewed down [110]. (b,c) Single layers of the different MnTe$_6$ octahedra viewed down [001], where (b) is an image of Mn1 layers and (c) a Mn2 layer. In all panels, one unit cell is outlined, MnTe$_6$ octahedra are shown, and arrows indicate relative alignments of Mn moments within the trigonal plane.}%
\label{Structure}%
\end{figure}

While its crystal structure is interesting, very little research has been reported on \MST.\cite{Rimet1981,Vincent1986}  The initial report of its existence and physical properties discussed the magnetism within the framework of an incorrect stoichiometry and structure,\cite{Rimet1981} and thus the initial hypothesis of ferrimagnetism could not be fully evaluated.  Despite containing nominally Mn$^{2+}$ with no orbital moment  (3$d^5$ with S=5/2, L=0), the magnetism displays a large anistropy field on the order of 10\,T at 5\,K.\cite{Rimet1981}  The crystal structure was reported several years after the initial characterization,\cite{Vincent1986} but the magnetic properties were not revisited.  \MST has a trigonal crystal structure (space group No. 163) that is shown in Fig.\,\ref{Structure}.\cite{Vincent1986}  Fig.\,\ref{Structure} also contains the magnetic structure obtained from our neutron diffraction data.  \MST is composed of MnTe$_6$ octahedra that are edge-sharing within the $ab$ plane (Mn1 site), and along with Si-Si dimers this creates layers of Mn$_2$Si$_2$Te$_6$.  The layered framework is analogous to that of CrSiTe$_3$, which is hexagonal and has a van der Waals gap between the layers.  In \MST, however, the layers are linked by the filling of one-third of the octahedral holes within the van der Waals gap by Mn atoms at the Mn2 site, yielding a composition of \MST.\cite{Vincent1986}  Importantly, the multiplicity of Mn1 is twice that of Mn2.   

In this work, \MST was characterized using magnetization, specific heat and electrical resistivity measurements, as well as via powder x-ray and single crystal neutron diffraction.  We find \MST to be a ferrimagnet due to antiparallel alignment of moments on the Mn1 and Mn2 atomic positions, the different multiplicity of which yields a net magnetization with moments preferring to lie in the trigonal plane.   The data, including diffuse neutron scattering, reveal the existence of strong spin correlations well-above \TC, which may be associated with short range order or the persistence of correlated excitations in the paramagnetic region.  The experimental probes were complemented by first principles calculations that revealed a competition between antiferromagnetic (AFM) exchange interactions that are frustrated with respect to each other.  Interestingly, the longer-range, third-nearest neighbor coupling (Mn1-Mn2 pairs at 5.41\AA) dominates over the in-plane coupling of second-nearest neighbors at 4.06\AA\, (Mn1-Mn1 pairs).  The net result is an antiferromagnetic configuration close in energy to the ferrimagnetic one, and this competition for the ground state leads to a relative suppression of the ordering temperature.

\section{Experimental Details}

\MST single crystals were obtained by melting a stoichiometric mixture of the elements in a vacuum sealed quartz ampoule at $T$=1125\,K.  High-purity Si lump (Alfa Aesar 99.9999\%), and Te shot (Alfa Aesar, 99.999\%) were combined with Mn granules (99.98\% Alfa Aesar) that were arc-melted prior to use.  Single crystals of \MST were mechanically isolated from the as-grown ingot.

Field-cooled magnetization measurements were performed in a Quantum Design Magnetic Property Measurement System.  AC magnetic susceptibility, electrical transport, and specific heat capacity measurements were performed in a Quantum Design Physical Property Measurement System.  Powder x-ray diffraction measurements were performed using a PANalytical X'Pert Pro MPD with a Cu K$_{\alpha,1}$ incident-beam monochromator.  An Oxford PheniX Cryostat was utilized to obtain data between 20\,K and 300\,K.  Additional powder x-ray diffraction measurements were performed between 300 and 400\,K on a PANalytical X'Pert diffractometer with Cu K$_{\alpha}$ radiation and an XRK900 oven-type furnace. The furnace chamber was purged with He for 12 hours prior to initiating the high-temperature measurements and gas was continuously flowed throughout the measurements. The measurements were delayed at each temperature for 30 minutes to allow the sample to equilibrate. At 300\,K, Rietveld refinement of data from the low $T$ stage yielded $a$=7.0321(6)\AA\, and $c$=14.249(1)\AA, while refinement of the data from the high $T$ stage yielded $a$ =7.0343(2)\AA\, and $c$ =14.249(2)\AA.

Single crystal neutron diffraction data were measured using the HB3A diffractometer at the High Flux Isotope Reactor at Oak Ridge National Laboratory (ORNL). The neutron wavelength of 1.546\,\AA \, from the bent perfect Si-220 monochromator was used for the data collections,\cite{chakoumakos2011four} and 380 reflections were measured at 4, 100, and 380\,K.  The nuclear-only refinement at 380\,K yielded occupancies equal to unity within the error bars, and thus the occupancies were fixed at unity for all $T$.  The magnetic refinement at 4\,K required three equivalent magnetic domains.  All refinements were performed using the program FullProf.\cite{FullProf}

Diffuse neutron scattering measurements were performed on the CORELLI diffuse scattering spectrometer at the Spallation Neutron Source located at ORNL.  CORELLI is a time-of-flight spectrometer with a pseudo-statistical chopper, which generates both an elastic scattering signal (with an average energy resolution $\approx$1\,meV) and a total scattering signal from a single measurement.  An approximately 7\,mg single crystal was attached to an Al plate in a manner to facilitate inspection of the HHL scattering plane. The measurements were taken at temperatures of 6, 120, and 350\,K, where the sample was rotated in steps of 3$^{\circ}$ over ranges of 120$^{\circ}$ (6\,K), 120$^{\circ}$ (120\,K) and 90$^{\circ}$ (350\,K). A significant amount of time was spent at 120\,K measuring 30 degrees centered around the 002 Bragg reflection.  Data around 002 were also collected on warming from 6\,K to 120\,K. Mantid was utilized to perform the Lorentz and spectrum corrections and merge the full volume of the scattering data.\cite{Michels2016} 

First principles calculations were performed using the linearized augmented plane-wave (LAPW) code WIEN2K,\cite{wien} within the generalized gradient approximation.\cite{perdew}  Spin-orbit coupling was only included during the calculation of magnetic anisotropy within the observed ground state. Sphere radii of 2.04, 2.5 and 2.5 Bohr were chosen for Si, Mn and Te, respectively, with an RK$_{max}$ of 7.0 (here RK$_{max}$ is the product of the smallest sphere radius and the largest planewave expansion wavevector).  All calculations used the experimental, room-temperature lattice parameters of $a$=7.029\AA\, and $c$=14.255\AA\, taken from Ref.\,\citenum{Vincent1986}.  For each magnetic configuration, the internal coordinates were relaxed within the magnetic ordering pattern until the forces were less than 2\,mRyd/Bohr.  Optimization within the magnetic state yields significantly different atomic coordinates than optimization within the non-magnetic state (an energy gain of some 250\,meV/Mn results), suggesting a strong coupling of magnetism to the lattice.  The structure theoretically optimized within the ferrimagnetic ground state is much closer to the experimental structure than that optimized within the non-magnetic state.  For example, the value for Mn1 of $\Delta$ z $\equiv$ z$_{Exp.}$ - z$_{Calc}$ is -8.7 $\times$ $10^{-4}$ for the ferrimagnetic ground state but nearly 5 times as large in magnitude at 4.02 $\times$ 10$^{-3}$ for the non-magnetic state.  Similarly large ratios apply for the Si z and Te y and z coordinates, with the Te x coordinate $\Delta$x roughly twice as large for the non-magnetic state as for the ferrimagnetic state.  Additional details are provided in the Supplemental Information.

\section{Results and Discussion}

\subsection{Magnetization, neutron diffraction and theoretical calculations}

\begin{figure}[t!]%
\includegraphics[width=\columnwidth]{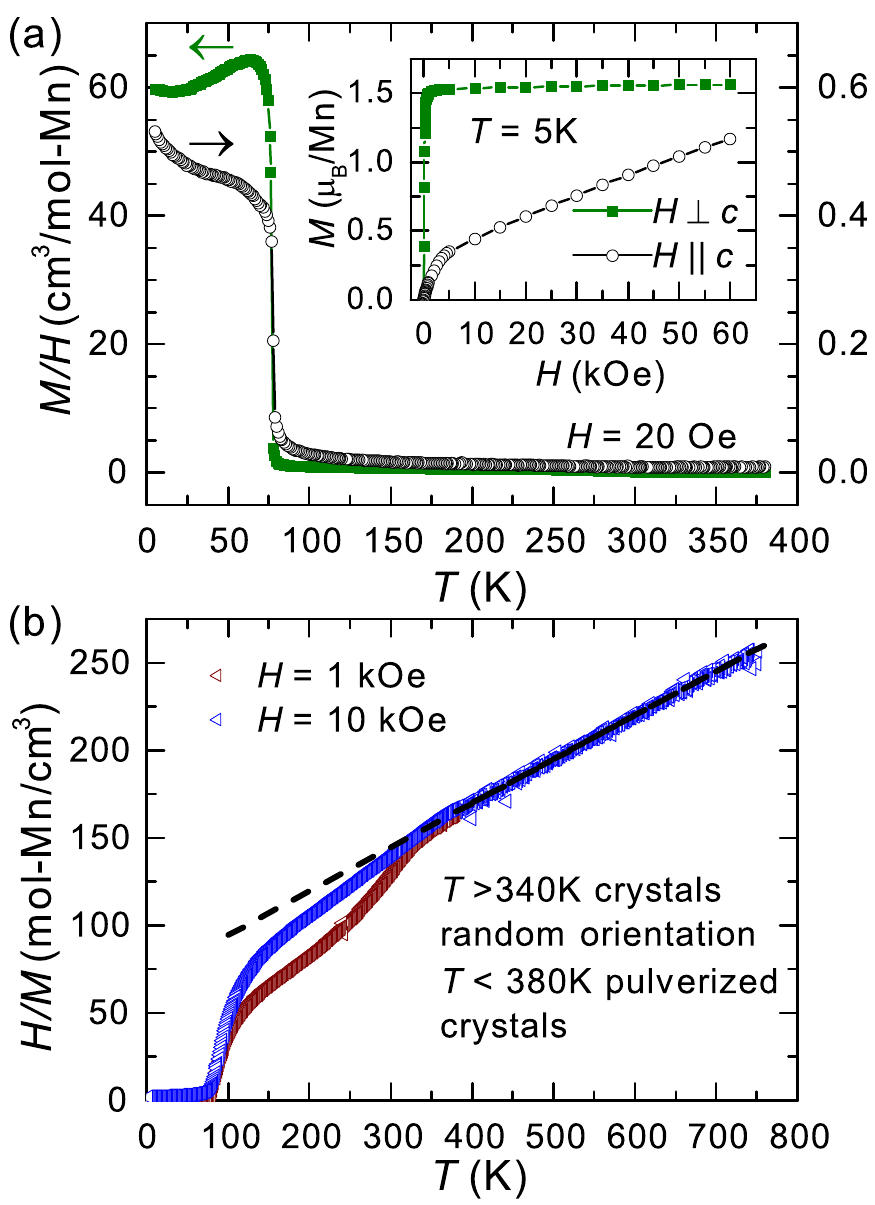} \\%
\caption{(a) Anisotropic magnetization data for \MST upon cooling in an applied field of 20\,Oe; the left axis ($H$ $\perp c$) and right axis ($H$ $\parallel c$) have the same units.  The inset shows isothermal magnetization data at $T$=5\,K, and together these results reveal easy-plane magnetization. (b) Inverse susceptibility data combining high and low $T$ measurements.  The dashed curve represents a Curie-Weiss fit extended to lower $T$.}%
\label{Mag}%
\end{figure}

\begin{figure}[t!]%
\includegraphics[width=\columnwidth]{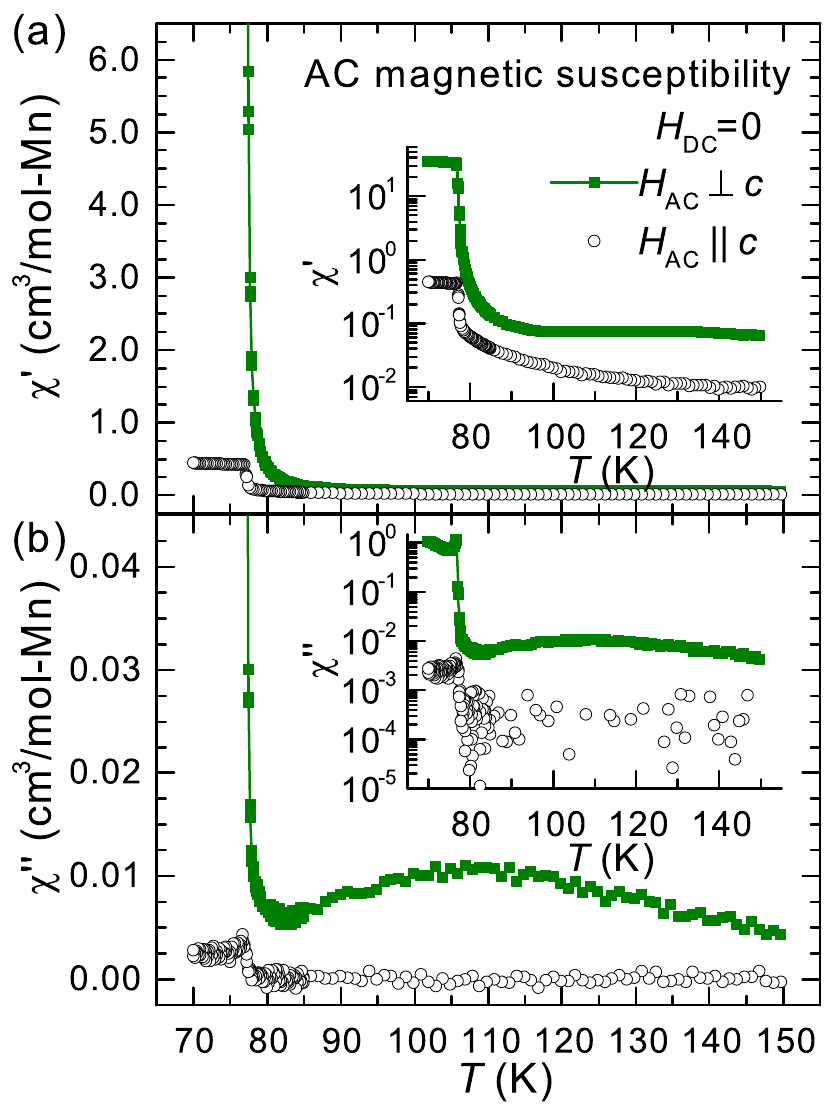} \\%
\caption{(a) The real component $\chi$' and  (b) the imaginary component $\chi$'' of the AC magnetic susceptibility data near the ferrimagnetic ordering temperature. The insets show the data on log-scale using the same units as the primary panels.  Data collected upon cooling using $A$=2\,Oe and $f$=997\,Hz.}%
\label{ACMag}%
\end{figure}

The magnetization $M$ data for single crystalline \MST are shown in Fig.\,\ref{Mag}a.  \MST is observed to have a Curie temperature of \TC=78\,K.  The ordering temperature was also confirmed by AC magnetic susceptibility measurements, the results of which are shown in Fig.\,\ref{ACMag}.  The anisotropy of $M$ below \TC suggests the ordered moments lie primarily within the $ab$-plane.  This is demonstrated by a significantly larger $M$ when the applied field $H$ lies within the $ab$-plane as compared to when $H || c$; note the different vertical axes in the main panel of Fig.\,\ref{Mag}a.  Consistent with this, below \TC the magnetization saturates rapidly for $H || ab$ and reaches $\approx$1.6\,$\mu_B$/Mn at $T$=5\,K (inset Fig.\,\ref{Mag}a).  The data imply an anisotropy field of approximately 9\,T (90\,kOe), and a small ferromagnetic component is observed for $H || c$.  These data are consistent with the initial report on \MST.\cite{Rimet1981}  We note there is no remanant moment for either orientation, which is different from the prior report and, to some extent, suggests the present single crystals are of high quality.

Fig.\,\ref{Mag}b plots the magnetization data as $H/M$, which is equivalent to $1/\chi$ when the susceptibility can be defined as $\chi=M/H$ (when $M$ is linear in $H$). The data demonstrate Curie-Weiss behavior between approximately 350 and 750\,K, the region where 1/$\chi$ is linear in $T$.   The data above 400\,K were fit to a simple Curie-Weiss law, $\chi$= C/($T-\Theta$), where C is the Curie constant and $\Theta$ the Weiss temperature.  This fitting produced an effective moment of 5.6$\mu_B$/Mn and a Weiss temperature of -277\,K.  The effective moment is consistent with the presence of Mn$^{2+}$ ions and the negative Weiss temperature indicates antiferromagnetic correlations.  The saturation magnetization (inset, Fig.\,\ref{Mag}a) is about one-third of that expected on the basis of this effective moment, which suggests the presence of one uncompensated Mn$^{2+}$ local moment per formula unit in the ordered phase.   In Fig.\,\ref{Mag}b, the low-temperature data were collected on a large mass of ground crystals, while high-temperature data were collected using six small crystals sealed under vacuum in a thin quartz tube (the crystals are free to rotate in an applied field).

The AC susceptibility $\chi_{AC}$ = d$M$/d$H$ can be described by in-phase (real) $\chi$' and out-of-phase (imaginary) $\chi$'' components, which are shown in Fig.\,\ref{ACMag}.   The out-of-phase component $\chi$'' relates to dissipative losses, for instance the movement of domain walls in ferromagnets; $\chi$'' is typically zero for simple antiferromagnets and non-zero in ferromagnets and sometimes in metals.\cite{Balanda2013}  The AC data shown in Fig.\,\ref{ACMag} were collected in zero applied DC field, with a small amplitude $A$=2\,Oe and a frequency $f$=997\,Hz.  These data reveal a sharp onset at the Curie temperature, with $\chi$' and $\chi$'' increasing below \TC for both orientations.  In addition, the data reveal anisotropy above \TC, which was also observed with the DC measurements. A comparison of $\chi$' and the DC analog $M/H$ ($H$=20\,Oe) is shown in the Supplemental Information.

\begin{figure}[t!]%
\includegraphics[width=\columnwidth]{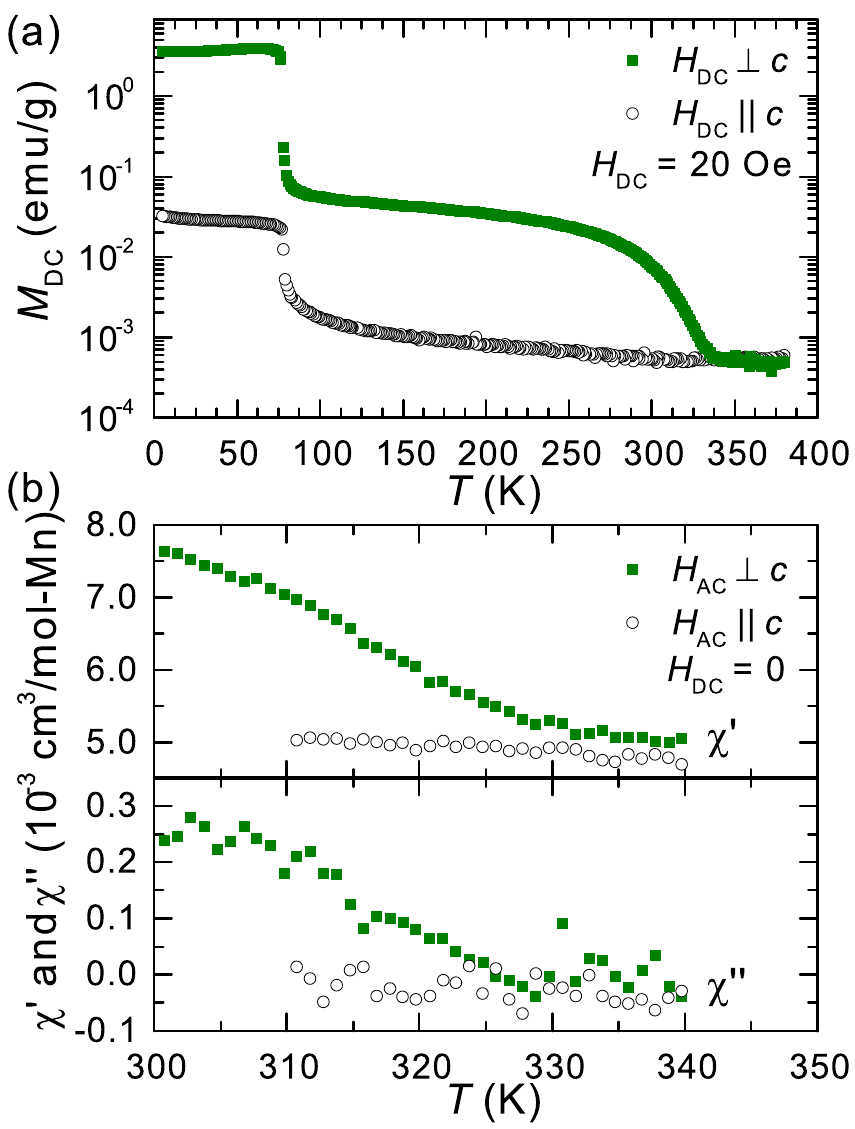} \\%
\caption{(a) Anisotropic DC magnetization data collected upon cooling in an applied field of 20\,Oe.  (b) Anisotropic AC susceptibility data showing the real $\chi$' and imaginary $\chi$'' components.  The AC data were collected using $A$=14\,Oe and $f$=997\,Hz with zero applied DC field.}%
\label{Short2}%
\end{figure}

As shown in Fig.\,\ref{Mag}b, the low-field magnetization data on ground crystals deviate from the high $T$ Curie-Weiss behavior below $\approx$ 330\,K. Single crystal magnetization data reported in Fig.\,\ref{Short2} demonstrate that this behavior is associated with a relatively abrupt onset of magnetic anisotropy below $\approx$ 330\,K, which is observed in both the low-field DC data (Fig.\,\ref{Short2}a) and the AC data (Fig.\,\ref{Short2}b).  The anisotropy is the same as that in the ferrimagnetic phase below \TC (easy-plane anisotropy).  Upon cooling below 330\,K, $\chi$'' becomes non-zero for in-plane data, and it reaches a maximum near 110\,K before rising sharply at \TC.  This may suggest the anisotropy is associated with, or derived from, a ferromagnetic component (uncompensated moment) that is present for $H \perp c$.  Indeed, isothermal magnetization data reveal a small (soft) ferromagnetic contribution for $H \perp c$ in the region \TC $< T <$330\,K, which increases upon cooling.  The low-field $M(H)$ data at $T$=110\,K are shown in Fig.\,\ref{MH110}, and the corresponding AC data are shown in the Supplemental along with $M(H)$ curves at different $T$.  Interestingly, $\chi$' plateaus near 85-110\,K, which would suggest that this ferromagnetic component has saturated.  That is, this plateau in $\chi$' seems to suggest that the anisotropy does not originate in precursory short-range order of the ferrimagnetic phase, which would be expected to continue increasing all the way down to \TC as the correlation length begins to diverge.  However, the ferromagnetic-like contribution suggests the anisotropy originates from some form of magnetic order, as opposed to a crystal-field induced anisotropy of a paramagnet. As discussed below, our diffraction measurements did not reveal any significant structural change near 330\,K, but a structural contribution and/or origin cannot be ruled out at this time.  The true nature of the transition near 330\,K is thus unclear and warrants further investigation.  As discussed in the Supplemental Information, this behavior seems to be intrinsic to our single crystals.

\begin{figure}[t!]%
\includegraphics[width=\columnwidth]{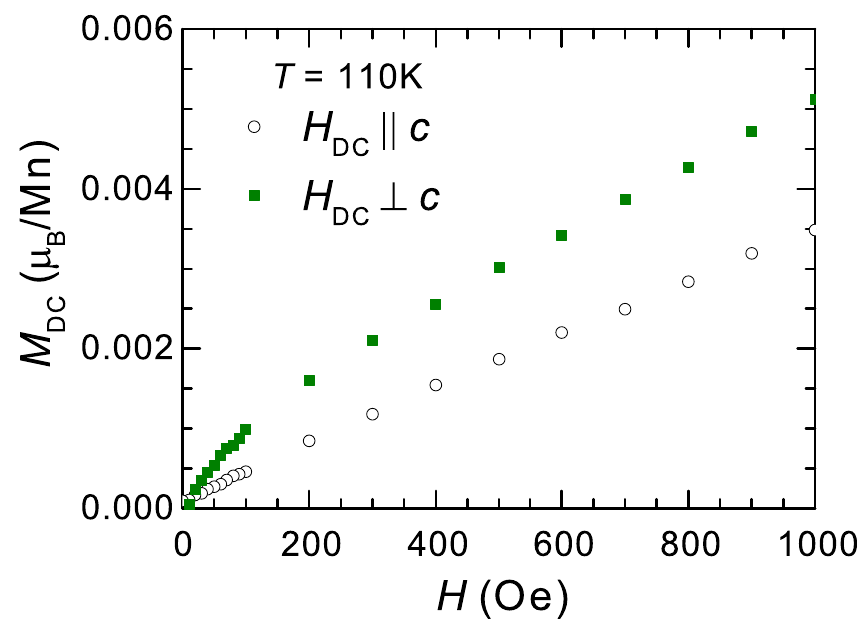} \\%
\caption{Isothermal DC magnetization data at 110\,K demonstrating the ferromagnetic contribution for $H \perp c$ in \MST.}%
\label{MH110}%
\end{figure}

\begin{figure}[ht!]%
\includegraphics[width=\columnwidth]{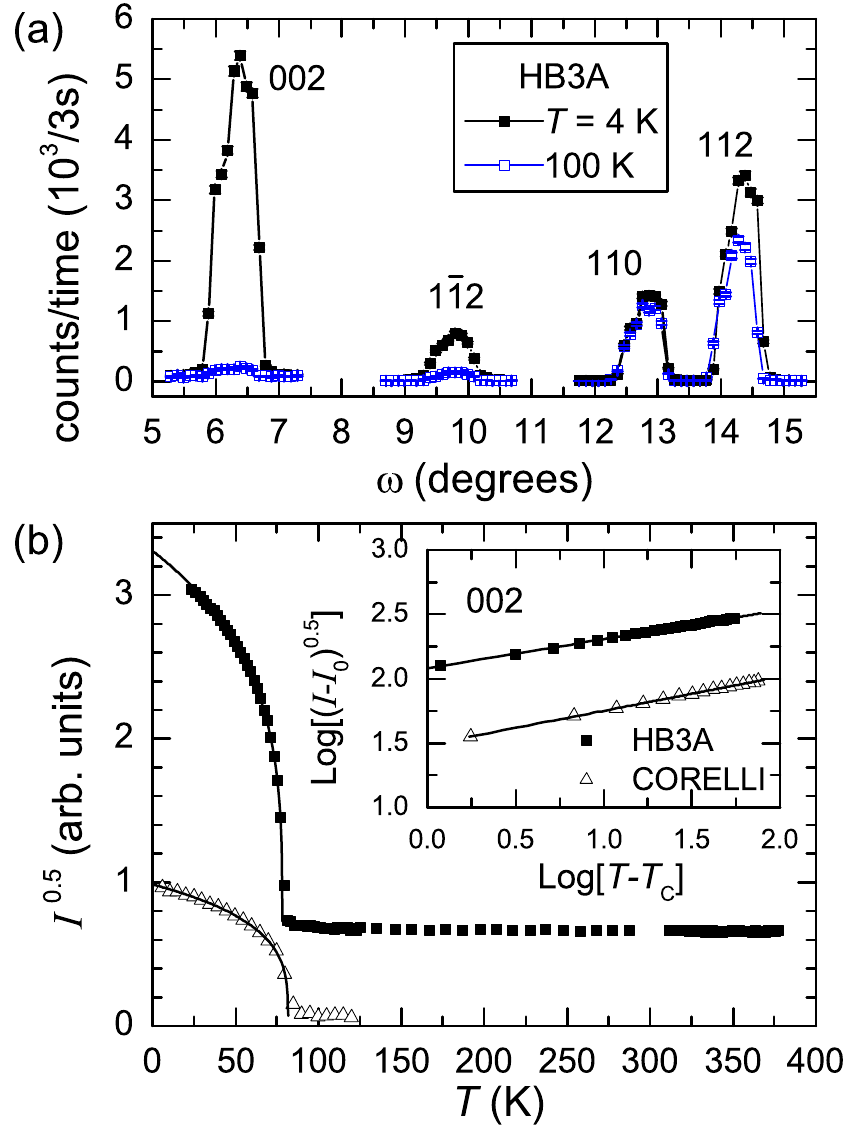} \\%
\caption{(a) Rocking curves of various Bragg reflections at 4 and 100\,K.  The increase of intensity upon cooling into the magnetically-ordered state is strongest for reflections with non-zero L. (b) Order parameter for the ferrimagnetic structure (002 intensity), with data from the single crystal neutron diffractometer (HB3A) and the elastic diffuse neutron scattering spectrometer (CORELLI).  The solid curves are fits to a simple power law, and the inset shows the data and fits in a log-log plot.}%
\label{ND1}%
\end{figure}

Single crystal neutron diffraction and spectroscopy were utilized to examine the nuclear and magnetic structures of \MST.  As shown in Fig.\,\ref{ND1}, the intensity of the 002 Bragg reflection increases upon cooling through \TC. This suggests a strong in-plane component of the moment, consistent with the magnetization data.  The lack of additional scattering at the 110 reflections at low $T$ is consistent with the moment being in the $ab$-plane, but this observation alone is not definitive given the presence of three magnetic domains.   The square root of the integrated magnetic intensity is proportional to the ordered moment, and can thus be used as a magnetic order parameter.  Order parameter data employing intensity from the 002 reflection were obtained using both HB3A and CORELLI, as shown in Fig.\,\ref{ND1}b where fits to power law behavior are included.  The data in Fig.\,\ref{ND1}b from CORELLI are the integrated intensity of 002, while that from HB3A were constant Q data collections; neither data set are corrected for background.  Additional temperature-dependent data from HB3A are shown in the Supplemental Information.

The order parameter fits shown as solid curves in Fig.\,\ref{ND1}b yield a critical exponent $\beta$ $\approx$ 0.25 and \TC values consistent with that obtained from magnetization measurements, though some minor discrepancies are observed because the data were collected while ramping temperature in different conditions. The fits employ a simple power law, $(I-I_0)^{0.5}$=$A$(\TC-$T$)$^{\beta}$, where $I_0$ represents the non-magnetic intensity (background and nuclear, assumed temperature independent).  Data from just below \TC down to $\approx$40\,K are fit, and the resulting curves are extended to lower $T$ in Fig.\,\ref{ND1}b for comparison.  For the CORELLI data we obtain $\beta$=0.26(1) and \TC=81.7(2)\,K while for the HB3A data we obtain $\beta$=0.23(1) and \TC=78.5(1)\,K.  The standard deviations originate from the fitting procedures, but the absolute errors on critical exponents estimated in this way are expected to be larger.  The inset contains these same results in log-log representation to emphasize the similar temperature-dependence for the two data sets, and to highlight the good quality of the fits.  The critical exponent of $\beta$ $\approx$ 0.25 for \MST can be compared to those obtained for simple mean field models.  For instance, a 2D Ising system is expected to have $\beta$=0.125, while the 3D Ising and Heisenberg models have $\beta$=0.326 and 0.367, respectively.\cite{yeomans1992statistical}  In CrSiTe$_3$, a critical exponent of  $\beta$=0.12 was reported, suggesting 2D character of the magnetism in that van der Waals bonded ferromagnet.\cite{Carteaux1995,William2015}  \MST is a crystallographically three-dimensional material.  The deviation of $\beta$ from the simple 3D models is not surprising, however, given the existence of multiple Mn sites and ferrimagnetic order.

Refinement of the 4\,K neutron diffraction data from HB3A yielded the magnetic structure shown in Fig.\,\ref{Structure} ($R_f=4.6$ and $\chi^2=3.6$).  This spin structure is defined by parallel alignment of the moments on Mn1 combined with anti-parallel alignment to the moments on Mn2.  Within the limits of these data, the refined moments are the same on both Mn sites, consistent with the expectation of similar oxidation states (both positions are octahedrally-coordinated by Te).  At 4\,K, the refined moment is 4.2(1)$\mu_B$/Mn when the moments are constrained to be equal and to lie within the trigonal plane.  When permitted, a small moment of 0.2(2)$\mu_B$/Mn along the $c$ axis is observed, but the quality of the refinement does not improve.  However, the isothermal magnetization data $M(H)$ at 5\,K suggest a soft ferromagnetic component for $H || c$ (inset Fig.\,\ref{Mag}a), and the low-field $M(T)$ data are anisotropic below \TC (Fig.\,\ref{Mag}a).  It is possible that an applied field causes alignment of canted moments or domains, and the lack of an applied field during neutron diffraction leads to an average canting that is negligible. Note that in Fig.\,\ref{Structure}, the moments are shown to be along the  $a$ axis, but their specific orientation within the $ab$-plane cannot be determined due to symmetry.

In trying to understand this magnetic configuration, we first consider the crystal structure and apparent exchange interactions.  Mn1-Mn2 have an interatomic distance of 3.55\,\AA\ at room temperature,\cite{Vincent1986} and are linked through face-sharing octahedra along the $c$ axis (Fig.\ref{Structure}), which would lead to a direct exchange interaction that we expect to be antiferromagnetic (AFM) for these Mn$^{2+}$ ions.  The Mn1-Mn1 interaction is more complicated because it occurs via edge sharing octahedra, which results in a competition of direct interaction (AFM) and $\approx$\,90$^{\circ}$ Mn1-Te-Mn1 interactions that can be either ferromagnetic (FM) or AFM depending on the $p$ and $d$ orbitals involved.\cite{KhomskiiBook2014} Since the ground state is ferrimagnetic with parallel alignment of Mn1 moments, it is tempting to suggest that FM superexchange is the dominant in-plane interaction.  Such an explanation would seem consistent with the observation of ferromagnetism in the structurally-related material CrSiTe$_3$, in which the analogous edge-sharing octahedra have ferromagnetic coupling and orovide the shortest exchange pathway (there is no inter-plane occupancy).\cite{William2015}   Theoretical calculations were performed to get a better understanding of the coupling mechanisms.

First principles calculations were performed to understand the ferrimagnetic ground state and the associated \TC.  Energies of the following six states were calculated: non-magnetic, ferromagnetic (FM), ferrimagnetic states FI1 and FI2, and antiferromagnetic states AF1 an AF2.  FI1 is the ferrimagnetic ground state, as also observed experimentally, and its magnetic symmetry is shown in Fig.\,\ref{Structure}.  The first competing state is AF1, in which all Mn1-Mn1 planar neighbors are anti-aligned, while all Mn1-Mn2 nearest neighbors are aligned.  Additional details regarding the various magnetic configurations are in the Supplemental Information.  The energies associated with the magnetic states studied fall more than an eV per Mn below the non-magnetic state, and the Mn moments can accurately be considered local moments.

\begin{table}[h!]
\caption{The calculated relative energies for several magnetic ordering patterns, along with net magnetic moments reported as $\mu_B$/Mn.  In all cases, each Mn is calculated to have a moment consistent with Mn$^{2+}$.\label{Table1}}
\begin{center}
\begin{tabular}{|c|c|c|}
\hline
Ordering &  $\Delta$E  (meV/Mn) & M ($\mu_{B}$/Mn) \\ \hline
FI1 (ground state) & 0 & 1.67 \\ \hline
AF1 & 19.1 & 0 \\ \hline
FI2  &  31.5 & 1.67 \\ \hline
AF2 &  32.2 & 0 \\ \hline
FM & 105.4  & 4.82 \\ \hline
\end{tabular}
\end{center}
\end{table}

The calculations predict that the ferrimagnetic configuration is the ground state, being substantially lower in energy than the paramagnetic or ferromagnetic states (105\,meV/Mn below the ferromagnetic state).  As summarized in Table \ref{Table1}, AF1 is only 21\,meV/Mn above the ground state, while AF2 and FI2 are slightly more than 30\,meV/Mn above FI1.  The energy of the ground state relative to competing states determines \TC, which is to say that the existence of competing magnetic configurations acts to suppress \TC.  Using a mean field approach, the Curie temperature is one-third of the energy difference between the ground state and the competing state, which in this case gives 19.1\,meV/3\kB = 73.9\,K, in good agreement with the experimentally observed \TC=78\,K.  Ultimately, the similar energies of the competing states reveal a competition between different coupling mechanisms and/or pathways.

The theoretical analysis was extended by mapping the observed energetics to a Heisenberg model assuming an Ising spin S=$\pm$1, which yielded the first three effective exchange constants.  The definitions and relative values of these J$_{i}$ are given in Table \,\ref{Table}.  Compared to the third-nearest neighbor at 5.42\AA , the other Mn-Mn distances are $\approx$7\AA \,\, or greater and have been neglected.  The mean field approach to a Weiss temperature is one-third of the average interaction energy per magnetic ion, which in this case is $\Theta$ $\approx$-757\,K/3 = -252\,K (accounting for the multiplicities), which is in good agreement with the experimental value of -277\,K.  To give a sense for the energy scales, the calculations yielded J$_{1}$=-35(4)\,meV/Mn.
 
The calculations revealed that the first three exchange constants are all antiferromagnetic in nature, which results in a frustrated system of competing interactions.  As shown in Table\,\ref{Table}, J$_{1}$ provides the strongest exchange, and its AFM nature is consistent with our expectations based on the direct exchange pathway. Interestingly, the second-nearest neighbor interaction (J$_{2}$) is AFM and is weaker than the third-nearest neighbor interaction (J$_{3}$).  Thus, the longer-range J$_{3}$ dominates the in-plane J$_{2}$ and together with the strong J$_{1}$ produces FM alignment of the Mn1 moments within the $ab$-plane despite the energy cost due to an AFM J$_{2}$.   The occupancy of the Mn2 position is therefore essential for generating FM alignment of Mn1 moments in the ground state of \MST.  Thus, this seems to contrast the behavior observed in ferromagnetic CrSiTe$_3$ where the hypothetical Cr2 position is vacant but the in-plane coupling appears to be ferromagnetic.\cite{William2015}

The multiplicity of each exchange interaction is critical in determining the order of  competing ground states.  For instance, in ferromagnetic MnBi the nearest-neighbor exchanges are AFM in nature, but a larger number of longer-range, ferromagnetic interactions results in a high Curie temperature of 630\,K.\cite{Williams2016}  In \MST, the multiplicity of J$_{2}$ and J$_{3}$ are equal despite a different number of pertinent neighbors (due to crystallographic site multiplicities), and are a factor of three larger than that of J$_{1}$ (see Table \,\ref{Table}).  The energy minimization in the ground state comes from satisfying both J$_{1}$ and J$_{3}$.  However, due to the multiplicity difference, the first competing state (AF1) is realized by satisfying both J$_{3}$ and J$_{2}$.  Thus, due to the Mn-Mn coordinations, several competing states exist in \MST.

Since under some circumstances strong correlations can be present in 3$d$-based chalcogenides, we have also performed GGA+U calculations of the ferrimagnetic ground state and the ferromagnetic state, applying a Hubbard U of 3\,eV to the Mn $d$ orbitals.  There are two main effects of adding this U: the ferrimagnetic band gap is increased slightly, and the energetics of the system change significantly. In particular, the ferromagnetic state is now just 43\,meV/Mn above the ferrimagnetic state, to be compared with 105\,meV/Mn in the GGA-only  calculations. It is likely that similar effects would be obtained for the other excited states.  Since the GGA itself provides excellent agreement with both the magnitude of the Curie temperature and the Curie-Weiss $\Theta$ obtained from the exchange constants, it appears that the regular GGA is indeed an appropriate tool for studying this system.

Since \MST shows a surprisingly large anisotropy field - approximately 9\,T - we have also performed first principles calculations\cite{parkerUMn2Ge2,salesFe3Sn,shanavasMnBi} of the magnetic anisotropy in the ferrimagnetic state in this system, which necessitates the use of the spin-orbit interaction.  Here the moments are constrained to lie either along the $c$ axis, or within the $ab$ plane.  We find that the energy is some 0.65\,meV/Mn lower in the planar configuration, leading to a calculated anisotropy field $H_{A}$ of 13\,T, in reasonable agreement with the experimental observations.  It is interesting to note that this anisotropy appears to arise from an anisotropy in the Mn {\it orbital} moments, which are of magnitude 0.023\,$\mu_B$/Mn in the axial configuration (all orbital moments are parallel to the corresponding spin moments and hence are of opposite sign for Mn1 and Mn2), but are found to be 0.037\,$\mu_B$/Mn1 and -0.048\,$\mu_B$/Mn2 in the planar configuration.  The existence of an orbital moment implies a deviation from the 3$d^5$, S=5/2 electronic configuration, though it appears that only a small orbital contribution exists.

It is interesting to compare the magnetic exchange constants in \MST to those in MnTe.  MnTe has the hexagonal NiAs structure-type in its bulk form, and it contains the same exchange pathways as defined in Table \ref{Table}, but with slightly different bond angles, interatomic distances, and multiplicities.\cite{Szuszkiewicz2006}  MnTe has an antiferromagnetic structure composed of ferromagnetic planes that stack antiferromagnetially along the $c$ axis.\cite{Adachi1961}  The planar J$_2$ is ferromagnetic in MnTe,\cite{Szuszkiewicz2006} and thus the interactions in MnTe are not frustrated.  Therefore, while the magnetic structures of \MST and MnTe are similar (ferromagnetic planes stacked antiferromagnetically), the origins are rather different.  In both compounds, J$_2$ is associated with edge-sharing octahedra (see Table \ref{Table}). The octahedra are nearly ideal in MnTe (bond angles of 90.0$\pm$0.1$^{\circ}$), while they are rather distorted in \MST (for instance, one bond angle drops to 85.1$^{\circ}$).  In relation to J$_2$, the distortions in \MST lead to a significant deviation from the ideal angle of 90$^{\circ}$ (87.2$^{\circ}$) associated with Mn-Te-Mn superexchange, as well as a shorter interatomic distance for the coupling (4.06\,\AA\, compared to 4.13\,\AA\, in MnTe).  This comparison further highlights the importance of  frustrated interactions in \MST, which produce competing magnetic ground states.  As discussed above, the energy difference between the lowest energy (magnetic) configurations determines the magnetic ordering temperature, and thus the existence of competing interactions/states leads to a suppressed Curie temperature in ferrimagnetic \MST.  However, since the dominant exchange interactions in MnTe are not frustrated, they do not result in a competing ground state that lies close in energy.  The bulk ordering temperature of MnTe is thus rather large, \TN =310\,K.\cite{Adachi1961}

\begin{table}[h!]
\caption{Schematic defining the three nearest neighbor exchange constants.  All are found to be antiferromagnetic (J$<$0), and the values shown were obtained for an Ising spin of $\pm$ 1 in the Hamiltonian.  The relevant multiplicities and distances in the relaxed cell are provided; see Fig.\,\ref{Structure} for more crystallographic details.  The interaction multiplicity takes into account the number of neighbors at a given site and the associated crystallographic multiplicity (four Mn1 and two Mn2 per unit cell).}
  \label{Table}
  \centering
  \begin{tabular}{m{3cm} m{6cm}}
     \begin{minipage}{0.25\columnwidth}
    \includegraphics[width=1.3\columnwidth]{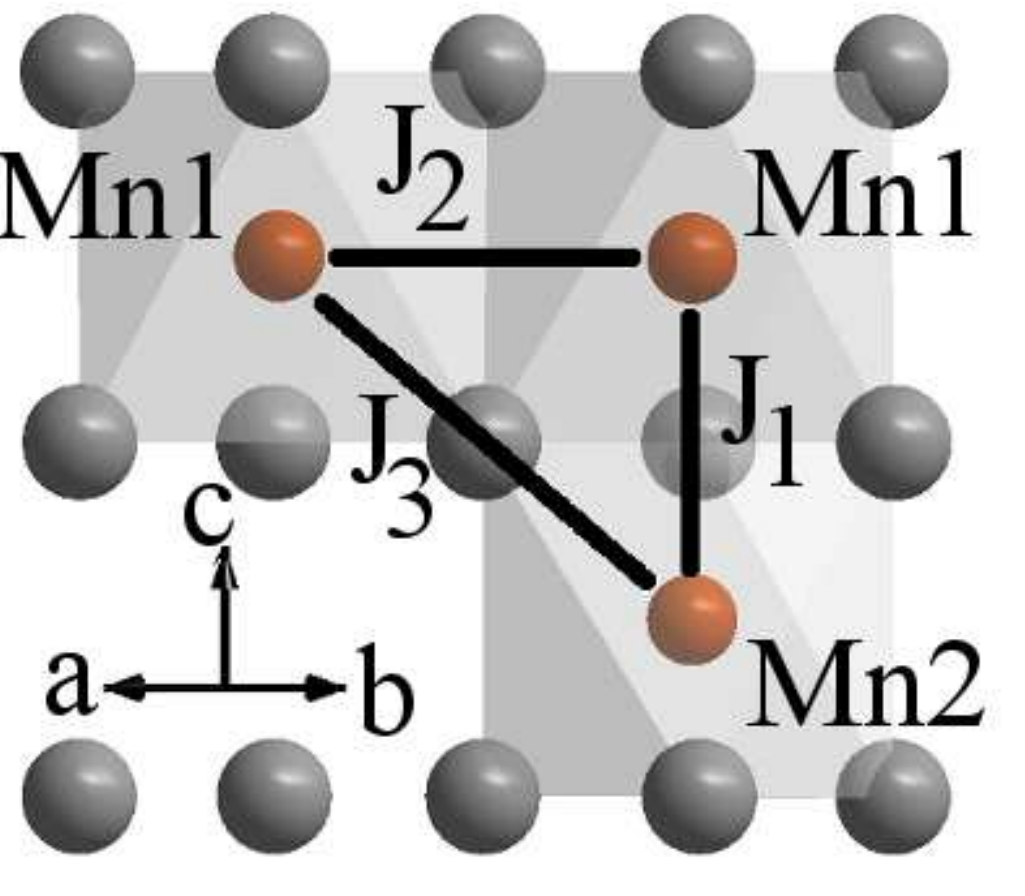}
    \end{minipage}
     &
     \begin{tabular} {ccc}
		Exchange  & Multiplicity & Distance \\
		 (K/Mn) & (per unit cell) & (\AA) \\
     \hline
      J$_{1}$ = -402(50) & 4 & 3.54\\
      J$_{2}$ = -73(9) & 12 & 4.06\\
      J$_{3}$ = -172(22) & 12 &  5.42 \\
     \end{tabular}
   \end{tabular}
\end{table}

\subsection{Lattice behavior and electrical resistivity}

\begin{figure}[h!]%
\includegraphics[width=\columnwidth]{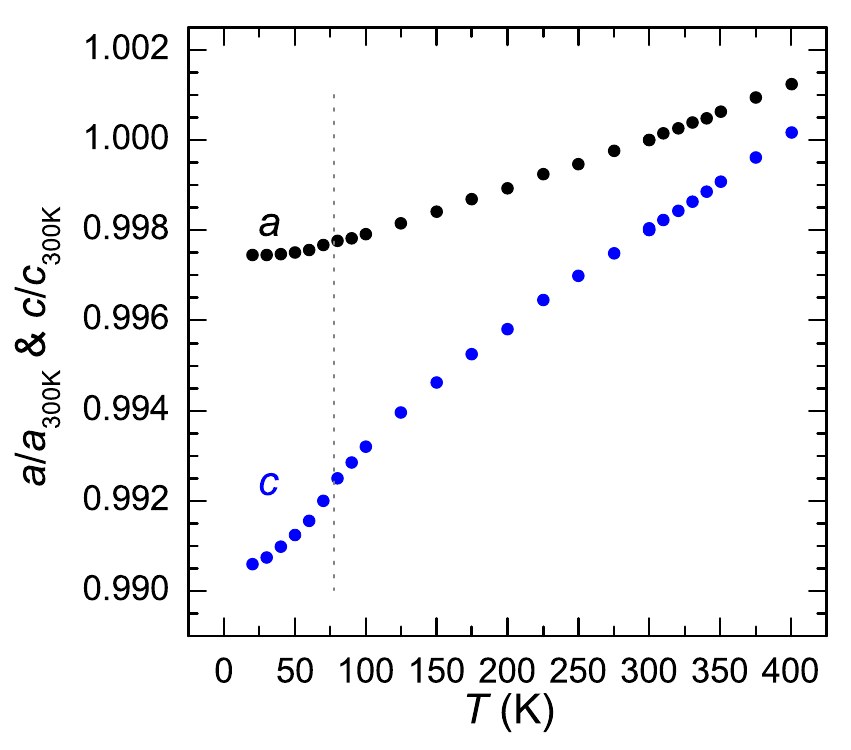}%
\caption{Temperature dependence of the lattice parameters normalized to values at 300\,K, with data for $c$ offset for clarity.  Error bars are smaller than the size of the data markers, and the dashed line indicates \TC.}
\label{Lattice}%
\end{figure}

As shown in Fig.\,\ref{Lattice}, the evolution of the $a$ and $c$ lattice parameters appears smooth across \TC and near 330\,K.  The thermal expansion along the $a$ axis is fairly typical for all $T$.  The $c$ parameter, however, shows a change in temperature dependence across \TC and an unusual curvature up to near room temperature. This seems to bolster the idea that an ordering transition occurs near 330\,K and has some coupling to the lattice, though additional measurements would be useful.  The neutron diffraction data at 100 and 380\,K refined well within the published crystal structure.

\begin{figure}[h!]%
\includegraphics[width=\columnwidth]{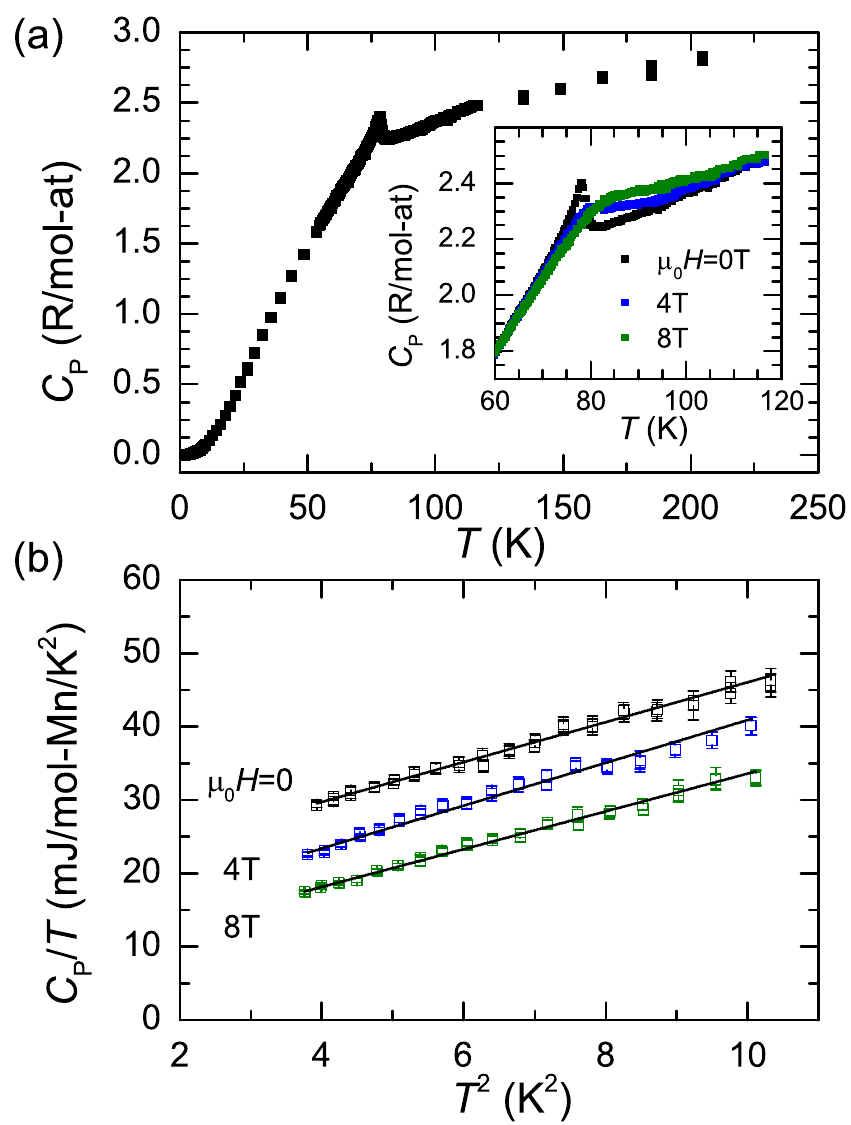}%
\caption{(a) Specific heat capacity of \MST, with inset showing the field dependence of the anomaly observed near \TC. (b) Field dependence of the the low-temperature specific heat data demonstrating a strong magnetic contribution.}%
\label{Cp}%
\end{figure}

The specific heat capacity of \MST is shown in Fig.\,\ref{Cp}.  An anomaly can be observed near \TC, and as shown in the inset an applied field ($H || c$) causes a broadening of this feature to higher temperature.  This behavior is consistent with ferromagnetic-like ordering, where an applied field suppresses fluctuations and broadens the specific heat anomaly by removing more of the magnetic entropy above \TC. While we do not have a good lattice (phonon) standard for \MST, an overestimation of the entropy change that occurs near \TC can be made by integrating $C_P/T$ with a straight line for a background (see Supplemental Information).  When fitting between 60 and 95\,K, this procedure yields an entropy change of only 8\% of that expected for ordering of S=5/2 local moments on Mn$^{2+}$.  The magnetic entropy per mole of Mn in the paramagnetic state is $S_{mag}$ = RLn[2S+1] = R Ln(6) = 14.9 J/mol-Mn/K, and the integration from 60 to 95\,K only yields $\Delta S_{mag}\approx$1.2\,J/mol-Mn/K.  Again, we emphasize that this is likely to be an overestimation of $\Delta S_{mag}$ due to the use of a linear background; this approach is utilized to emphasize how much entropy is released away from \TC (both above and below).  A similarly small $\Delta S_{mag}$ is observed in MnPS$_3$, where two-dimensional order is observed above \TC,\cite{Joy1992,Takano2004} as well as in CrSiTe$_3$ where short range correlations are observed above \TC and the magnetism is strongly coupled to the lattice.\cite{Casto2015,William2015}

As shown in Fig.\,\ref{Cp}b, there is a strong contribution to the specific heat capacity at low temperature, and this decreases with increasing applied field.   At the lowest temperatures, the data can be well-described by a Debye term ($\beta T^3$) plus a linear term ($\gamma T$), the latter of which is the typical way to include an electronic contribution.  In this case, however, \MST is a semiconductor and the strong magnetic-field dependence also points to a magnetic origin; $\gamma$ should not be associated with the typical electronic Sommerfeld coefficient.  As shown in the Supplemental Information, the low $T$ magnetic contribution in \MST (and CrSiTe$_3$) can also be described using a term proportional to $T^{1.5}$. Data below 2\,K would likely help in considering the nature of the low $T$ magnetic contribution in \MST.

\begin{figure}[h!]%
\includegraphics[width=\columnwidth]{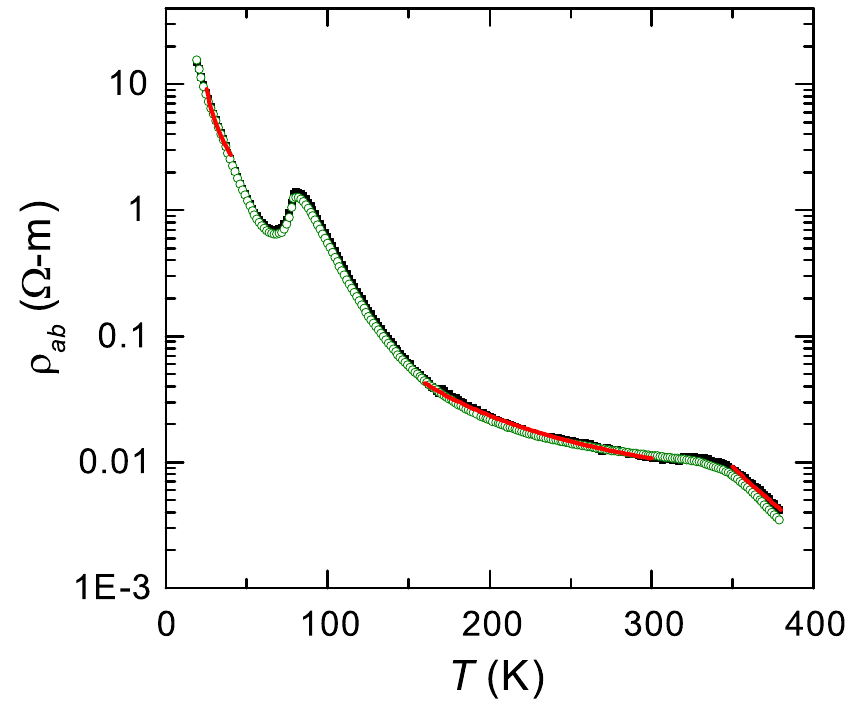}%
\caption{The in-plane electrical resistivity of \MST generally increases with decreasing temperature, with a notable feature near \TC.  The decrease of $\rho$ becomes more rapid above $\approx$330\,K.  The black and green curves are data collected on different crystals, and the red lines are fitted curves using $\rho$=$\rho_0$Exp[$E_A$/$k_BT$], which only fits the data over small temperature ranges but does reveal an apparent increase in activation energy upon warming across the two magnetic anomalies.}%
\label{Resist}%
\end{figure}

The in-plane electrical resistivity $\rho_{ab}$ of \MST is shown in Fig.\,\ref{Resist} for two crystals.  The value at 300\,K is consistent with that reported by Ref.\,\citenum{Rimet1981}, and similar to their data we observe activated-conduction and a strong anomaly at the ferrimagnetic ordering temperature.  An increase in the carrier mobility is expected below \TC due to the suppression of spin fluctuations, and this may explain the sharp dip in $\rho_{ab}$ below \TC.  The feature at \TC is somewhat dramatic, though, especially when considering the implications of a small $\Delta S_{mag}$ near \TC.  The present data, which extends to higher temperatures than previously reported, also shows a broad feature near 330\,K.  This appears to demonstrate the bulk-nature of magnetic ordering associated with the susceptibility anomaly observed at the same temperature in Fig.\,\ref{Mag}, and the associated behavior in Fig.\,\ref{Resist} could be related to carrier scattering or electronic structure effects.

To inspect the changes in $\rho(T)$ above and below the magnetic transitions, we fit the data to $\rho=\rho_0{\rm Exp[}E_A/k_B T{\rm]}$ where $E_A$ and $k_B$ are the activation energy and Boltzmann constant, respectively.  The fitted curves are shown in red in Fig.\,\ref{Resist}.  Firstly, we note that these fits suffer from inadequacy of the simple model (particularly at the lowest $T$), as well as the limited amount of data utilized.  However, the trends and magnitudes are worth noting. The activation energy obtained by fitting $\rho$ from 350-380\,K is roughly 0.3\,eV; this is the region where all of the material appears to be in a paramagnetic state. Below the magnetic anomaly observed at 330\,K, the activation energy decreases substantially, $E_A$=0.04\,eV for 160-300\,K.    Below \TC, the simple Arrhenius behavior is not a good model for the data.  For comparison sake, a small temperature range allows for a reasonable fit to be obtained, with $E_A$=0.007\,eV for 25-40\,K.  Interestingly, these results show the decrease of the activation energy with increasing magnetic order, and if $E_a$ is related to an energy gap then this is rather unusual behavior.  Note that the reduction of the activation energy at low $T$ could be due to the activation of low-level defects.

\subsection{Diffuse neutron scattering}

\begin{figure}[h!]%
\includegraphics[width=\columnwidth]{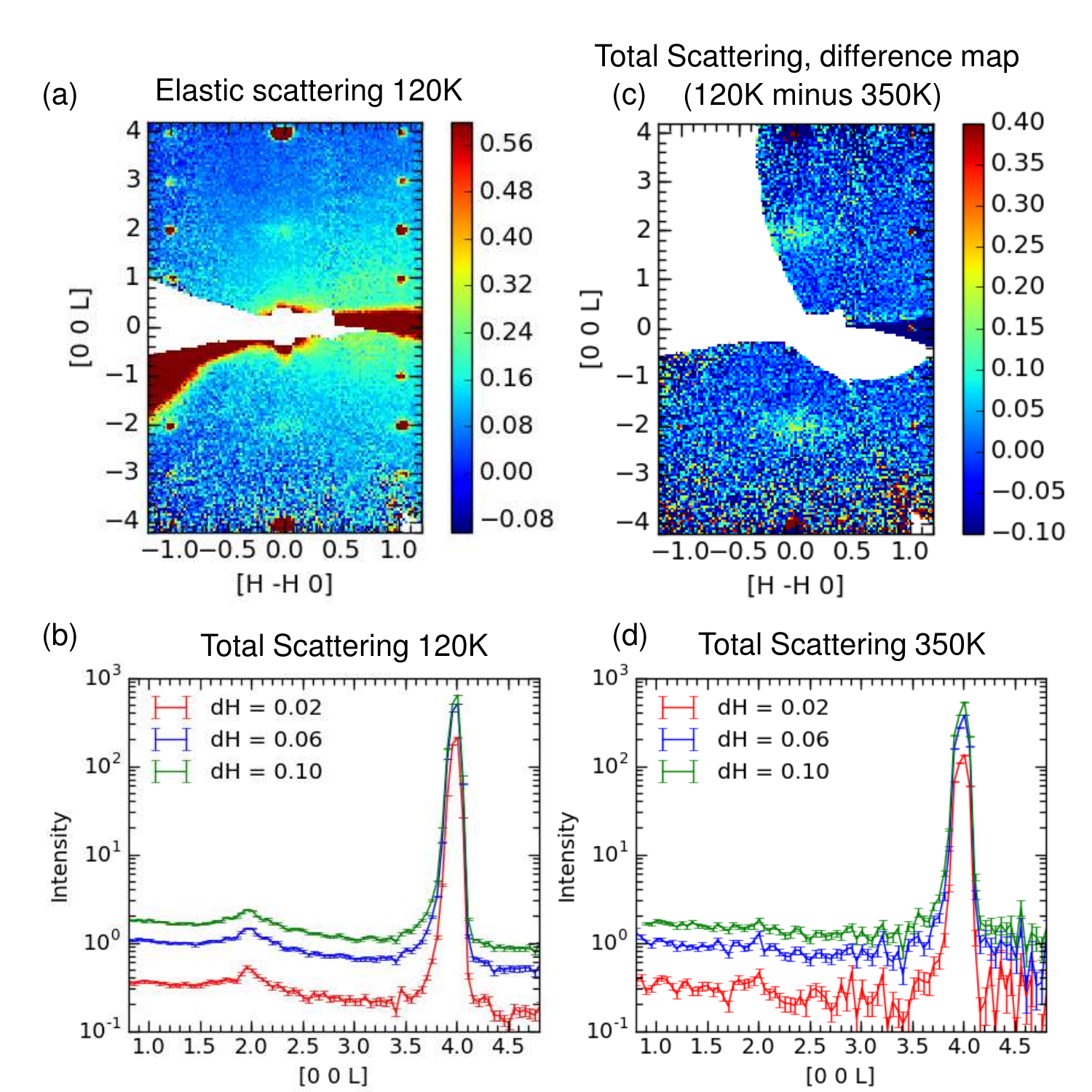}%
\caption{Data from the neutron elastic diffuse scattering spectrometer CORELLI. (a) Estimate of the elastic scattering at 120\,K showing diffuse scattering near the 002 Bragg reflections. Line scans of the total scattering at (b) 120\,K and (d) 350\,K across 002 and 004 reflections with different integration widths along H indicated in the legend.  (c) Difference map of total scattering (120\,K with 350\,K baseline) emphasizing the diffuse scattering near the 002 reflection.}%
\label{Diffuse120}%
\end{figure}

\begin{figure}[h!]%
\includegraphics[width=\columnwidth]{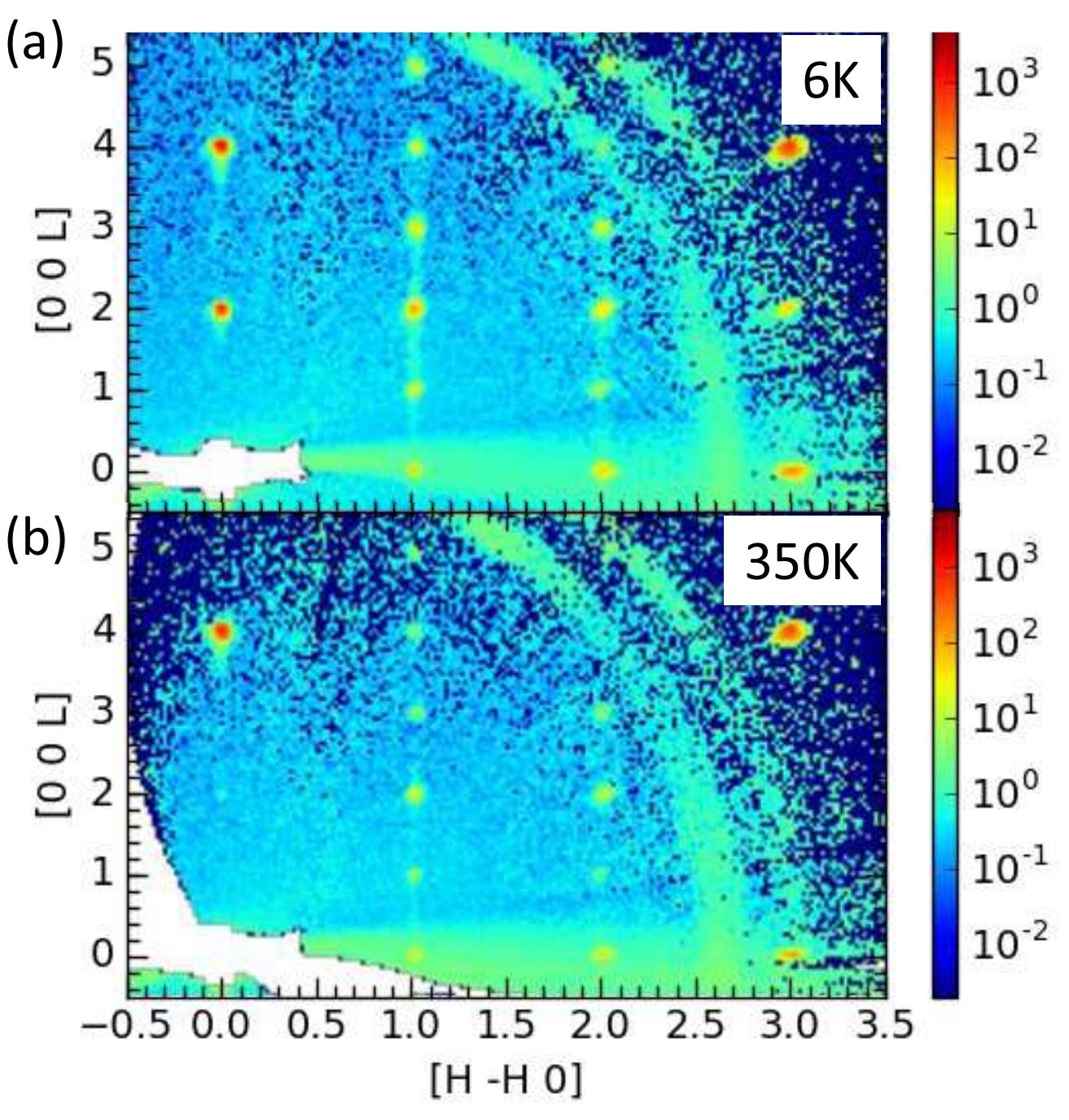}%
\caption{Total neutron scattering in the (H-HL) plane at (a) 6\,K and (b) 350\,K.  Diffuse scattering is observed along L, notably at 6\,K near 1-12.  These 2 D maps show the intensity from [-0.005 -0.005, 0] to [0.005, 0.005 0], i.e., a single bin, along the [H H 0] out-of-plane direction.}%
\label{Diffuse6}%
\end{figure}

Motivated by the bulk measurements that revealed anomalous behavior below 330\,K, neutron scattering was performed on CORELLI to search for diffuse scattering.  These measurements revealed diffuse scattering near the 002 reflection at 120\,K (Fig.\,\ref{Diffuse120}), and diffuse scattering along L in the (H-HL) plane at 6\,K (Fig.\,\ref{Diffuse6}).  The scattering data were binned into 3D bins with three orthogonal axes along [H H 0], [H -H 0] and [0 0 L] directions, respectively. The bin sizes are 1\%, 2\% and 5\% of the reciprocal vectors  [1 1 0], [1 -1 0] and [0 0 1], respectively. 

Figure \ref{Diffuse120}a is a 2D map of the elastic neutron scattering intensity for \MST at 120\,K, showing diffuse scattering around the 002 reflection.  The 2D map shows the integrated intensity from [-0.035 -0.035 0] to [0.035 0.035 0], i.e., 7 bins, along the [H H 0] out-of-plane direction. The strong scattering intensity reaching out from 000 in Fig.\,\ref{Diffuse120} was strongly angle dependent and did not appear consistent with the usual instrument background, but it was essentially independent of temperature.  We thus speculate that this feature may have been caused by the glue utilized to orient the small crystal.   

The diffuse scattering is emphasized in Fig.\,\ref{Diffuse120}c, which is a map of the intensity at 120\,K with the intensity at 350\,K subtracted. By contrast to the 002 reflection, the strong nuclear-only 004 reflection is sharp and is not surrounded by diffuse scattering.  This diffuse cloud around 002 is not observed at 6 or 350\,K, suggesting it is from short-range magnetic scattering.  It is interesting to speculate whether this diffuse scattering may be related to the weak ferromagnetic component of the magnetization seen below 330\,K. This hypothesis could be tested with additional temperature-dependent INS measurements.  The absence of this diffuse scattering at 6\,K suggests that it coalesces into the ferrimagnetic order.  This is consistent with the observation of diffuse scattering around the 002 reflection, which has a strong magnetic contribution in the ferrimagnetic phase.  The strong contribution from the long-range magnetic order to the intensity of the 002 reflection would also hinder detection of diffuse scattering at 6\,K.  

In Fig.\,\ref{Diffuse120}b, line scans along L are shown around the 002 and 004 reflections using different integration areas dH (resolution is inversely proportional to dH).  The data for 004 demonstrate the behavior expected for a nuclear-only Bragg peak; the 004 reflection has a sharp peak with little dependence on the temperature or integration area.  By contrast, the nuclear component of the 002 reflection is very weak (see Fig.\,\ref{Diffuse120}d) and the data at 120\,K demonstrate a very broad region of increased intensity centered around 002.  The sharpening of the intensity at 002 with decreasing dH (higher resolution) suggests the presence of a small, sharper Bragg reflection underneath a diffuse cloud of scattering.

The diffuse scattering in Fig.\,\ref{Diffuse120} demonstrate the existence of short-range spin correlations well-above \TC in \MST, though the nature of these correlations is unclear.  The scattering could be from short range order, potentially relating to the magnetization anomaly near 330\,K.  Alternatively, the diffuse scattering could be associated with excited states that persist above \TC and cause inelastic scattering that cannot be entirely excluded due to the nature of the experiment.  Above \TC, the paramagnons could be broad, un-gapped and without a defined dispersion, which would promote isotropic, diffuse scattering in these nearly-elastic measurements.   Both short-range order and excited states will exist above a second order magnetic transition, but in some cases these short-range correlations persist to much higher temperatures than is typically observed (alternatively, their correlations lengths are relatively longer above \TC).

The existence of short-range correlations above \TC is generally enhanced when \TC is suppressed relative to the strongest exchange interaction. Geometric frustration, competing interactions, or strong anisotropy in the exchange constants can all lead to such behavior.  In these cases, strong magnetic exchange interactions can promote short-range correlations (of the ground or excited states) well-above \TC because thermal energy has less influence near \TC.  As noted in Ref.\,\citenum{Jongh1974}, the relevant ratio to consider when thinking about short-range correlations near \TC is not $T/$\TC but rather $k_BT$/JS.  In many quasi-2D systems, anisotropic interactions lead to a relative suppression of  $k_B$\TC, and a strong JS can enhance correlation lengths above \TC. For \MST, competing antiferromagnetic interactions appear to provide the underlying mechanism that ultimately promotes the persistent, short-range correlations above \TC. A counter example is MnF$_2$, where the thermal energy is large relative to the exchange energy at the antiferromagnetic ordering \TN,\cite{Jongh1974} and as a result the majority of the magnetic entropy in MnF$_2$ is released below \TN $\approx$ 67\,K.\cite{Jongh1974,Boo1976}

At 6\,K, diffuse scattering was detected along L in the (H-HL) plane, as shown in Fig.\,\ref{Diffuse6}. The anisotropy of this low-temperature diffuse scattering suggests it is associated with a short correlation length along the $c$ axis.  The 6\,K diffuse scattering is centered around L=2, and is only observed around particular reflections (H-K = 3n+2) that are only contributed to by Mn scattering (both sites contribute).  This enhances the relative contribution of the magnetic intensity in these reflections and facilitates detection of this diffuse scattering.  Despite appearing to have a magnetic origin, there is perhaps a minor indication of this diffuse scattering at 350\,K (though this is within the noise of the background scattering).  Therefore, this feature may be associated with defects on the Mn positions.  From a magnetic perspective, the diffuse scattering seems to suggest that there is a short correlation length of spin canting along the $c$ axis.  While this seems plausible and consistent with our other measurements, the current data cannot determine the origin of this diffuse scattering.  An additional scattering experiment with an applied field would be the next step in determining the origin of this anisotropic diffuse scattering.  We note that the tails extending toward L = 0 (the asymmetry of 004 for instance) are an artifact of the time of flight measurement technique and are not observed in the elastic data (see Supplemental Information).

\section{Summary}

This work has characterized the magnetism of \MST and found it to be a ferrimagnet due to anti-parallel alignment of moments on different Mn atomic positions.  If the potential canting of the moments along the $c$ axis is ignored, then the magnetic structure is fairly simple.  By contrast, the mechanisms that lead to this ground state are relatively complex.  A competition between antiferromagnetic exchange interactions exists, and it is a third-nearest neighbor interaction that ultimately determines the ground state.  This frustration suppresses \TC and likely enhances the importance of fluctuations, leading to persistent short-range correlations above \TC.  Magnetization measurements also suggest that some additional ordering may exist below $\approx$330\,K.  Diffuse scattering was observed at 6\,K that may indicate short-range correlations of spin canting along the $c$ axis.  Due to the competing antiferromagnetic interactions, the manipulation of bond distances or the targeted substitution of the Mn2 atoms would likely yield new ground states with potentially higher ordering temperatures.

\section{Acknowledgments}

This work was supported by the U. S. Department of Energy, Office of Science, Basic Energy Sciences, Materials Sciences and Engineering Division. Work at ORNL’s High Flux Isotope reactor and Spallation Neutron Source were supported by the Scientific User Facilities Division, Office of Basic Energy Sciences, U.S. Department of Energy (DOE).


\renewcommand{\theequation}{S\arabic{equation}}
\renewcommand{\thefigure}{S\arabic{figure}}
\renewcommand{\thetable}{S\arabic{table}}

\setcounter{figure}{0}
\newpage
\section*{Supplementary Information}

This Supplemental Materials contains additional information regarding magnetization measurements and first principles calculations, as well as neutron diffraction and scattering data for Mn$_3$Si$_2$Te$_6$.  It also includes additional details for the analysis of specific heat capacity data, and a comparison of the low-temperature specific heat for Mn$_3$Si$_2$Te$_6$, CrSiTe$_3$ and CrI$_3$.

\section*{Magnetization measurements}

\begin{figure}[h!]%
\includegraphics[width=\columnwidth]{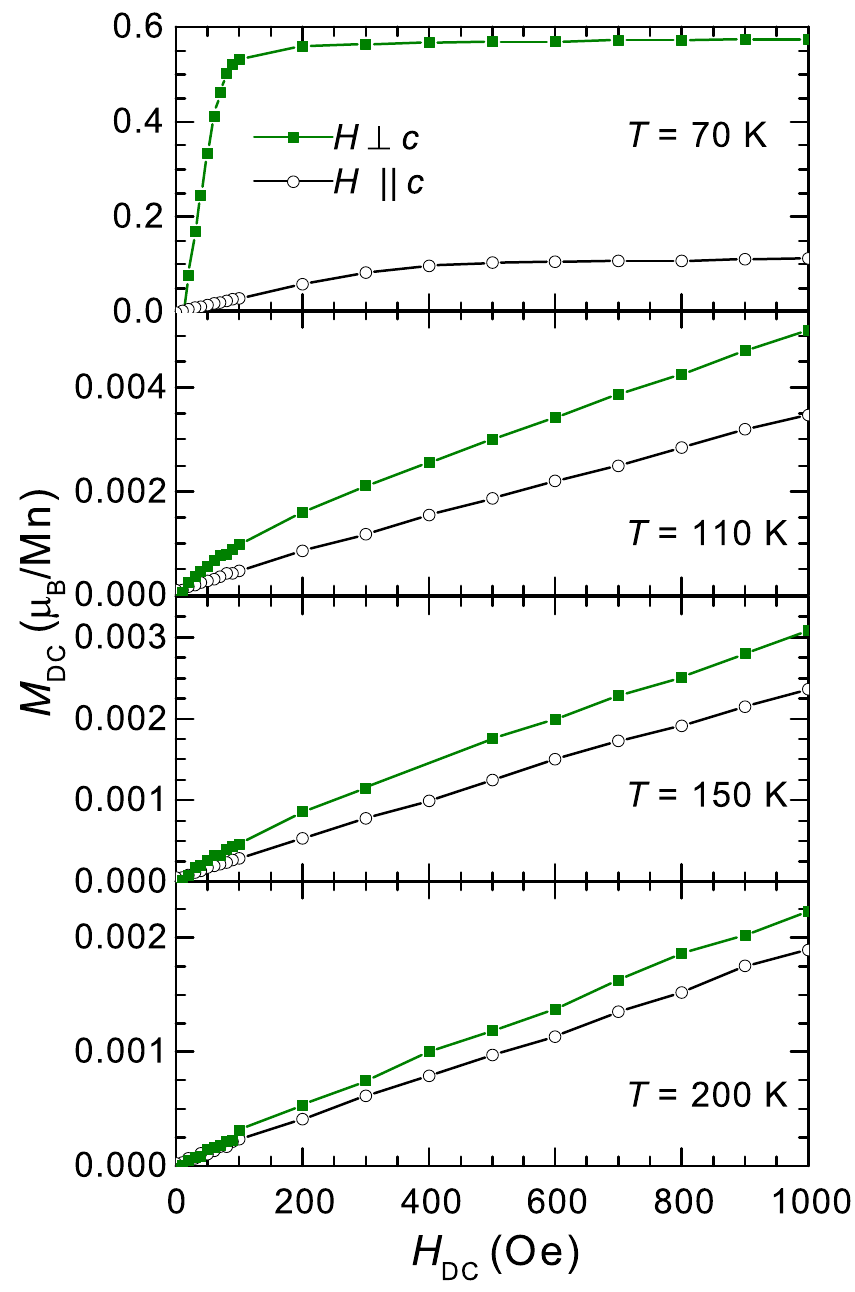}%
\caption{Isothermal magnetization curves showing the anisotropy of the induced moment as a function of applied field, with these plots highlighting the data at small applied fields.}%
\label{DCMH}%
\end{figure}

The bulk ordering temperature of $T_C$=78\,K was characterized using AC and DC magnetization measurements.  These measurements also revealed that the magnetization is anisotropic below 330\,K, which is well-above the bulk ordering temperature.  The anisotropy of the magnetization in this intermediate temperature region is consistent with that in the ordered region - the moment is largest for fields applied perpendicular to the $c$-axis.  This is emphasized in Fig.\,\ref{DCMH}, where isothermal magnetization data are shown for a variety of temperatures.  A small, ferromagnetic-like contribution for H $\perp$ $c$ can be seen above $T_C$=78\,K.  This is best observed at 110 and 150\,K, though  anisotropy is easily observed in the $M(H)$ data at higher $T$.

A small ferromagnetic component can sometimes be attributed to an impurity, but the observed anisotropy strongly suggests that this is intrinsic to our \MST single crystals.  Also, there was a change in the in-plane resistivity at approximately this same temperature.   None of the known binaries in the Mn-Si and Mn-Te phase diagrams order ferromagnetically as bulk materials and no other Mn-Si-Te ternary compounds are known.  Furthermore, we do not detect obvious signs of these binary impurities in powder x-ray diffraction or the 3D full volume single crystal scattering data from CORELLI.  We note that the previous work on \MST did not report data above 300\,K, nor did they report anisotropic susceptibility measurements, and the Curie Weiss fitting was performed using data collected in a large applied field.\cite{Rimet1981-SI}  It is worth mentioning that some of the potential impurities, such as MnSi$_{1.7}$,\cite{Zhou2007-SI} have complex magnetism that can be tuned to ferromagnetism using finite size effects.

\begin{figure}[h!]%
\includegraphics[width=\columnwidth]{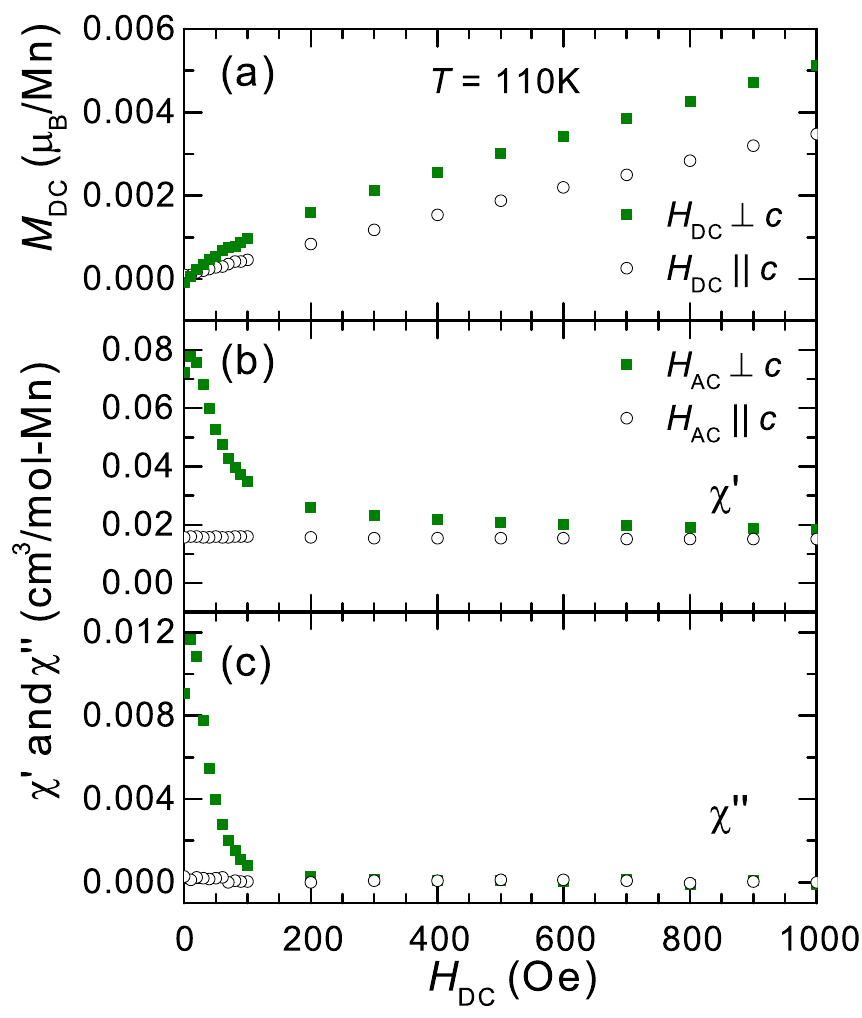}%
\caption{Isothermal magnetization for $T$=110\,K showing (a) DC magnetization and (b,c) AC susceptibility data.  The AC data were collected using an amplitude of $A$=14\,Oe and a frequency of $f$=997\,Hz.}%
\label{110K}%
\end{figure}

The AC magnetization measurement naturally probes the ferromagnetic-like contribution observed at low fields.  To demonstrate this, data at $T$=110\,K are examined in more detail in Fig.\,\ref{110K}.  The (soft) ferromagnetic contribution for $H$ $\perp$ $c$ is manifested as a larger $\chi$' than for $H$ $||$ $c$ (especially at $H_{DC}$=0).  When the DC field is increased from zero, the ferromagnetic component is rapidly quenched (or saturated), and thus $\chi$' decreases with increasing applied field for $H$ $\perp$ $c$.  Similarly, the out-of-phase component of the AC susceptibility ($\chi$'') decreases with applied field as the ferromagnetic portion is quenched (likely due to the loss of domain movement).  In the other orientation, $H$ $||$ $c$, $\chi$' is essentially independent of applied field because $M_{DC}$ is very linear with $H$ for this orientation (the ferromagnetic contribution is negligible above $T_C$=78\,K for $H$ $||$ $c$ , see Fig.\,\ref{DCMH}).  The anisotropy of $\chi$' and $\chi$'' are greatly reduced by $H$=1000\,Oe.

The data in Figures \ref{DCMH} and \ref{110K} thus reveal a ferromagnetic contribution that is related to the anisotropy in the magnetization below 330\,K.  This is consistent with the anisotropy observed in the magnetization measured during DC measurements, which is greatest at small applied fields.  

\begin{figure}[h!]%
\includegraphics[width=\columnwidth]{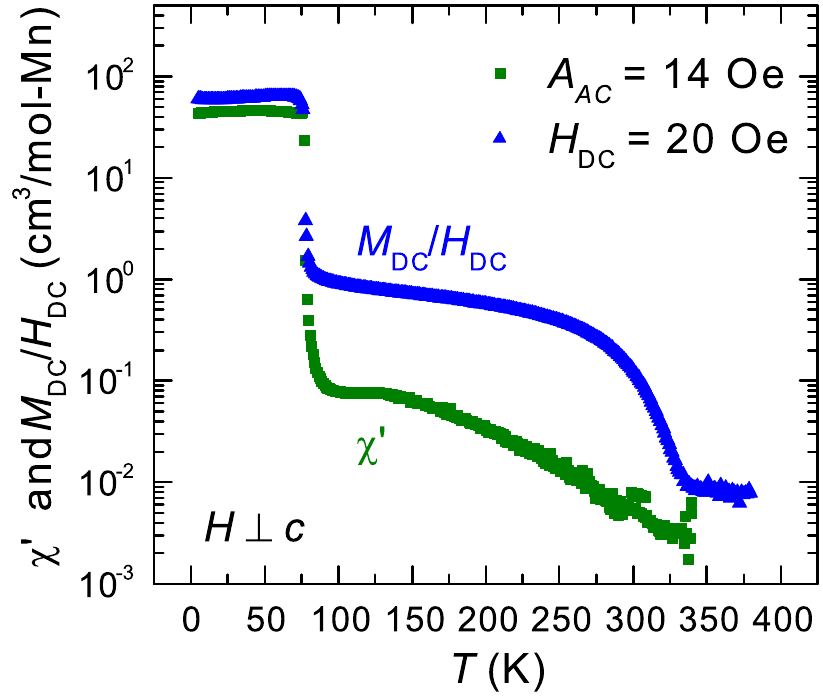}%
\caption{Comparison of AC susceptibility and quantity $M/H$ from DC measurements.  The DC measurements were conducted upon cooling in an applied field of 20\,Oe.  The AC data were obtained using $H_{DC}$=0, $A$=14\,Oe and a frequency of $f$=997\,Hz.  The data are for magnetization with fields lying in the basal plane of the trigonal crystal.}%
\label{ChiCompare2}%
\end{figure}

The small ferromagnetic contribution leads to a discrepancy between $\chi_{AC}$ and the DC analog $M_{DC}/H_{DC}$.  $M_{DC}/H_{DC}$ = $\chi$ when $M$ is linear in $H$ (as for a paramagnet at high $T$).  For comparison sake, we plot these two quantities together in Fig.\,\ref{ChiCompare2}.  The difference between these two quantities starts to grow below the 330\,K feature, and then increases again below $T_C$.  Note that for this orientation $H \perp c$, the demagnetization effects are expected to be small (field applied in the plane of a thin plate).

\section*{First principles calculations}

\begin{figure}[h!]%
\includegraphics[width=\columnwidth]{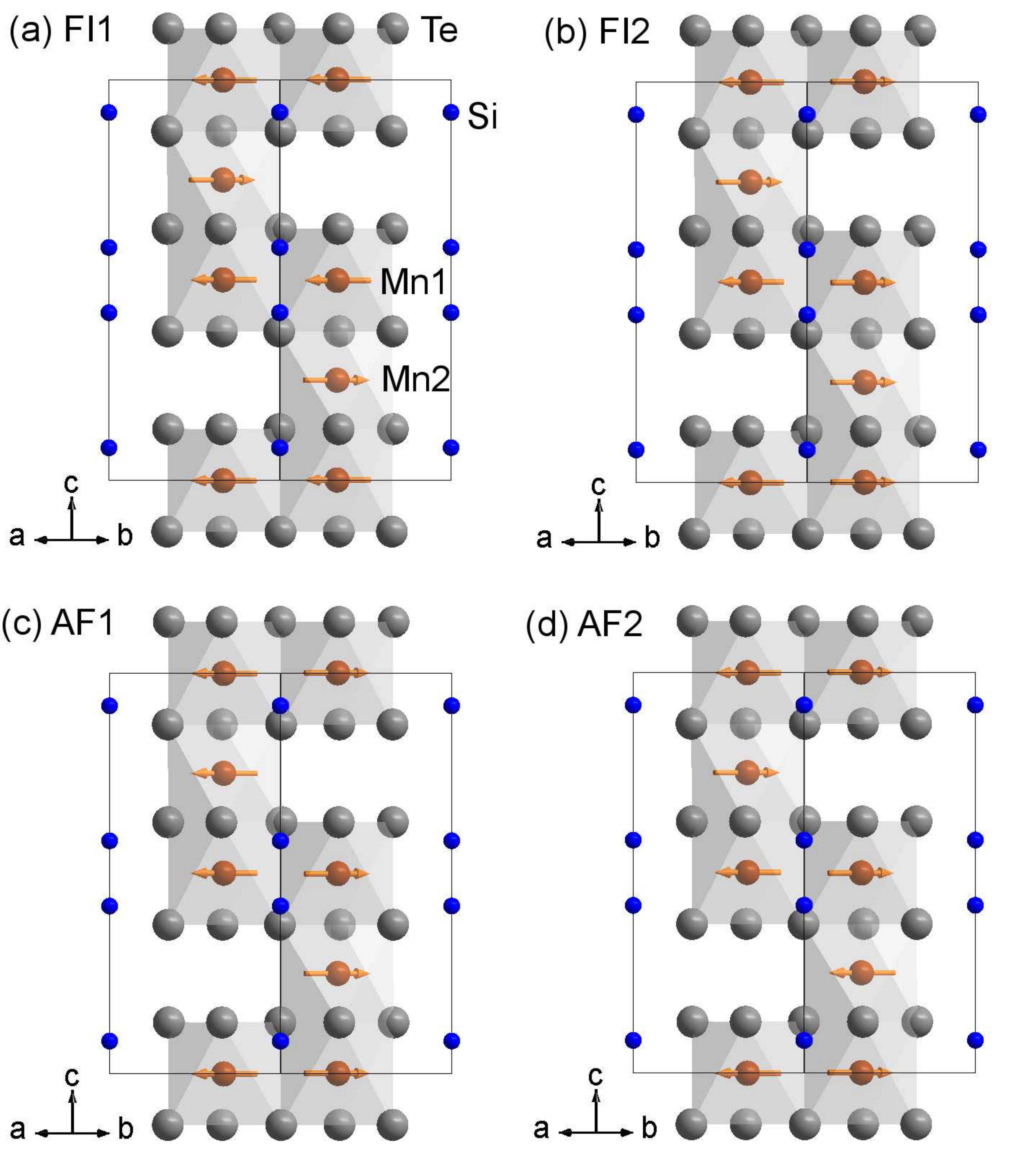}%
\caption{The magnetic configurations examined by first principles calculations in \MST are shown (the non-magnetic and ferromagnetic states are not shown). FI1 is the ground state, and relative to that energy we obtained AF1=20.8, FI2=32.6, and AF2=33.8\,meV/Mn.}%
\label{Excited}%
\end{figure}

Fig.\,\ref{Excited} shows the magnetic configurations explored with first principles calculations (the non-magnetic and ferromagnetic states are not shown).  Our first principles calculations included a non-magnetic state, a ferromagnetic state, the ferrimagnetic states FI1 (Mn1 and Mn2 spins opposed, the calculated ground state) and FI2 (Mn1-Mn1 planar neighbors anti-aligned and half of Mn1-Mn2 neighbors aligned and half antialigned), and two antiferromagnetic states AF1 and AF2 (both these states are 
$q=0$ states).  In state AF1, all Mn1-Mn1 planar neighbors are anti-aligned, while all Mn1-Mn2 nearest neighbors are ferromagnetically aligned.  In AF2 the Mn1-Mn1 planar neighbors are antialigned and Mn1-Mn2 nearest neighbors are anti-aligned.

The calculated energies were mapped to a Heisenberg model with J$_{1}$ the nearest-neighbor exchange (between Mn1-Mn2), J$_{2}$ the second-nearest neighbor exchange that is between planar Mn1-Mn1, and J$_{3}$ that couples Mn1 and a third-nearest spin at Mn2.  These calculations assumed an effective spin S of $\pm$ 1 in the Hamiltonian.  The values obtained are all antiferromagnetic: -34.6 , -6.3, and -14.8\,meV/Mn for J$_{1}$, J$_{2}$, and J$_{3}$, respectively.  Effective uncertainties of approximately 10-15 percent are obtained for each of these constants.  The uncertainty arises due to an over-determined set of equations determining the exchange constants (one more equation than variable, so that the constants are ``best-fit" constants).  Removal of one of the equations allows an exact set, for this equation subset, to be determined and thereby the uncertainty within the ``best-fit" constants assessed.

As mentioned, the first principles calculations depict a strong relationship of physical structure to the magnetism (specifically, the non-symmetry-dictated internal positions). In Table \ref{TabSI}, we compare the atomic coordinates extracted from the relaxations to the published structure.  Note that all calculations used the experimental lattice parameters of $a$ = 7.029\,\AA\, and $c$ = 14.255\,\AA, which were determined at room-temperature and reported in Ref.\citenum{Vincent1986-SI}.  LAPW sphere radii of 2.04 Bohr for Si,  and 2.5 for Mn and Te were employed and sufficient numbers of k-points - at least 600 in the full Brillouin zone - were used to determine energy differences.  An RK$_{max}$ of 7.0, where RK$_{max}$ is the product of the smallest LAPW sphere radius (in this case Si) and the largest planewave expansion wavevector, was employed.

In a certain sense, it is natural that signatures of the ferrimagnetic ground state would remain well above the Curie temperature of 78\,K.  The DFT results find the energy of the ferrimagnetic state to be well more than an eV per Mn below that of the non-magnetic state. This is more than two orders of magnitude larger than the 7\,meV/Mn scale of \TC, indicating a strong likelihood of ``disordered local moments" \cite{stocks-SI} well above \TC. Supporting this interpretation is the behavior of the resistivity, which shows semiconducting behavior well above \TC.  This is consistent with the semiconducting gap we find theoretically in the ferrimagnetic state, while the non-magnetic state from theory has a metallic behavior.   

\begin{table}[h!]
\caption{The internal coordinates of the non-symmetry dictated atomic positions.  We include the symmetry-dictated planar coordinates of Mn1 and Si for reference.  ``Exp." refers to the experimental published structure,\cite{Vincent1986-SI} NM to the non-magnetic calculations, and FI to the ferrimagnetic calculations. \label{TabSI}}
\begin{center}
\begin{tabular}{|c|c|c|c|c|c|}
\hline
Atom &  Site Symmetry & Origin &  x & y & z \\ \hline
Mn1  & 4f & Exp. & 1/3 & 2/3 & 0.00068\\ \hline
& &  FI-Calc. & 1/3 & 2/3 & 0.00155\\ \hline
 & & NM-Calc. &1/3 & 2/3 & 0.00470\\ \hline
  \\ \hline
Si & 4e & Exp &0 &0  & 0.08153\\ \hline
  & & FI-Calc. & 0 &0  & 0.08186\\ \hline
 & & NM-Calc. & 0 &0  & 0.08298\\ \hline
 \\ \hline
Te &  12i & Exp. & 0.65869 & 0.00361 & 0.12788 \\ \hline
&   & FI-Calc. & 0.65040 & 0.00081 & 0.12697 \\ \hline
&   & NM-Calc. & 0.64213 & 0.02033 & 0.11615 \\ \hline
\end{tabular}
\end{center}
\end{table}

\section*{Single crystal neutron diffraction data - HB3A data}

\begin{figure}[h!]%
\includegraphics[width=\columnwidth]{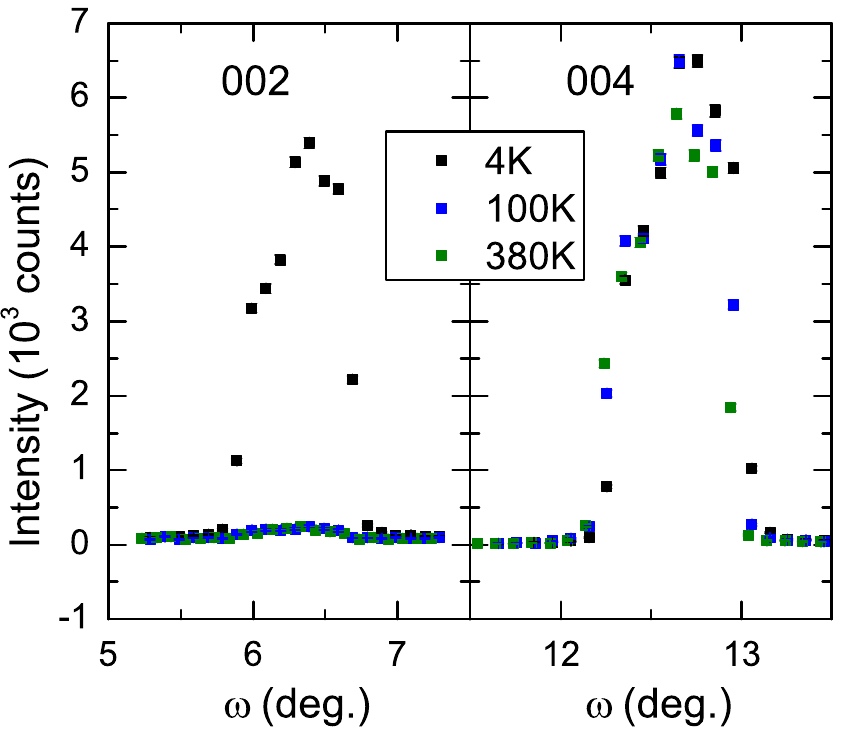}%
\caption{Rocking curves for single crystal neutron diffraction data of (a) the 002 Bragg reflection and (b) the 004 Bragg reflection at 4, 100, and 350\,K.}%
\label{Rocking}%
\end{figure}

\begin{figure}[h!]%
\includegraphics[width=\columnwidth]{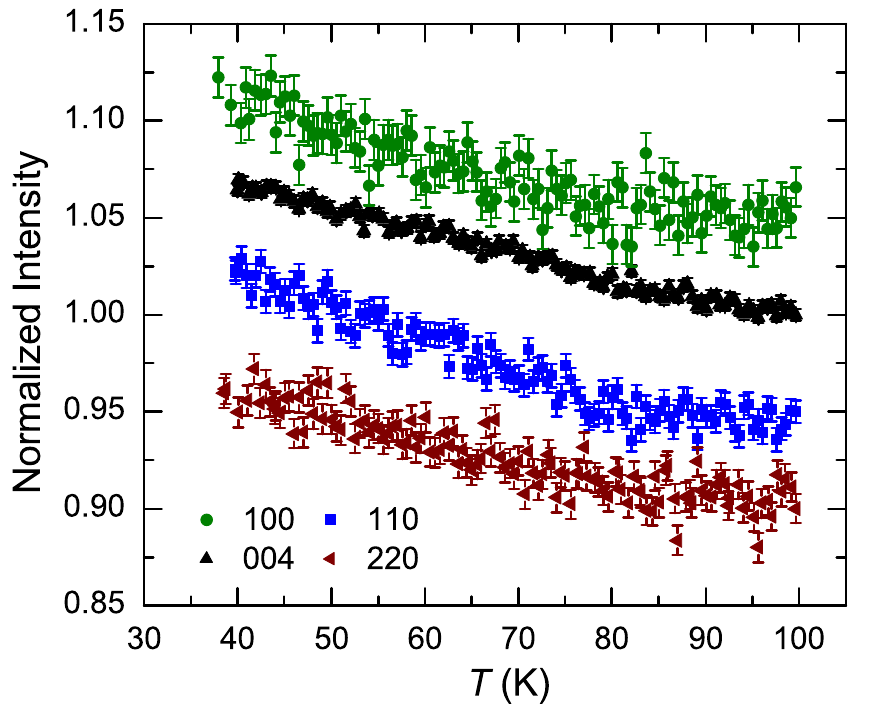}%
\caption{Temperature dependence of diffraction intensity at several Bragg reflections as obtained from the single crystal neutron diffraction beamline HB3A.  Based on the refined magnetic structure, the 004 reflection should not possess any magnetic contribution and thus the relative change with temperature is just a reflection of the evolution of the lattice or peak position (constant Q measurements).  The other reflections, with H $\neq$ 0, could contain some magnetic contribution if the moments have a component along the $c$ axis.  However, the relative change with $T$ for these peaks is similar to that for 004, suggesting that magnetic scattering is minor for the peaks shown here.   The data have been normalized and then offset from one another.}%
\label{NuclearSI}%
\end{figure}

Rocking curves from the four-circle diffractometer HB3A at the High Flux Isotope Reactor (ORNL) are shown in Fig.\,\ref{Rocking}.  These demonstrate a strong temperature dependence for the 002 reflection and not the 004 reflection.  The refined magnetic structure prohibits magnetic scattering contributions to (004), which is shown (Fig.\,\ref{NuclearSI}) to increase only slightly upon cooling across the magnetic ordering temperature of \TC $\approx$78\,K. The intensities of other nuclear-dominated reflections are shown in Fig.\,\ref{NuclearSI}.

\section*{Diffuse Scattering - CORELLI data}

\begin{figure}[h!]%
\includegraphics[width=\columnwidth]{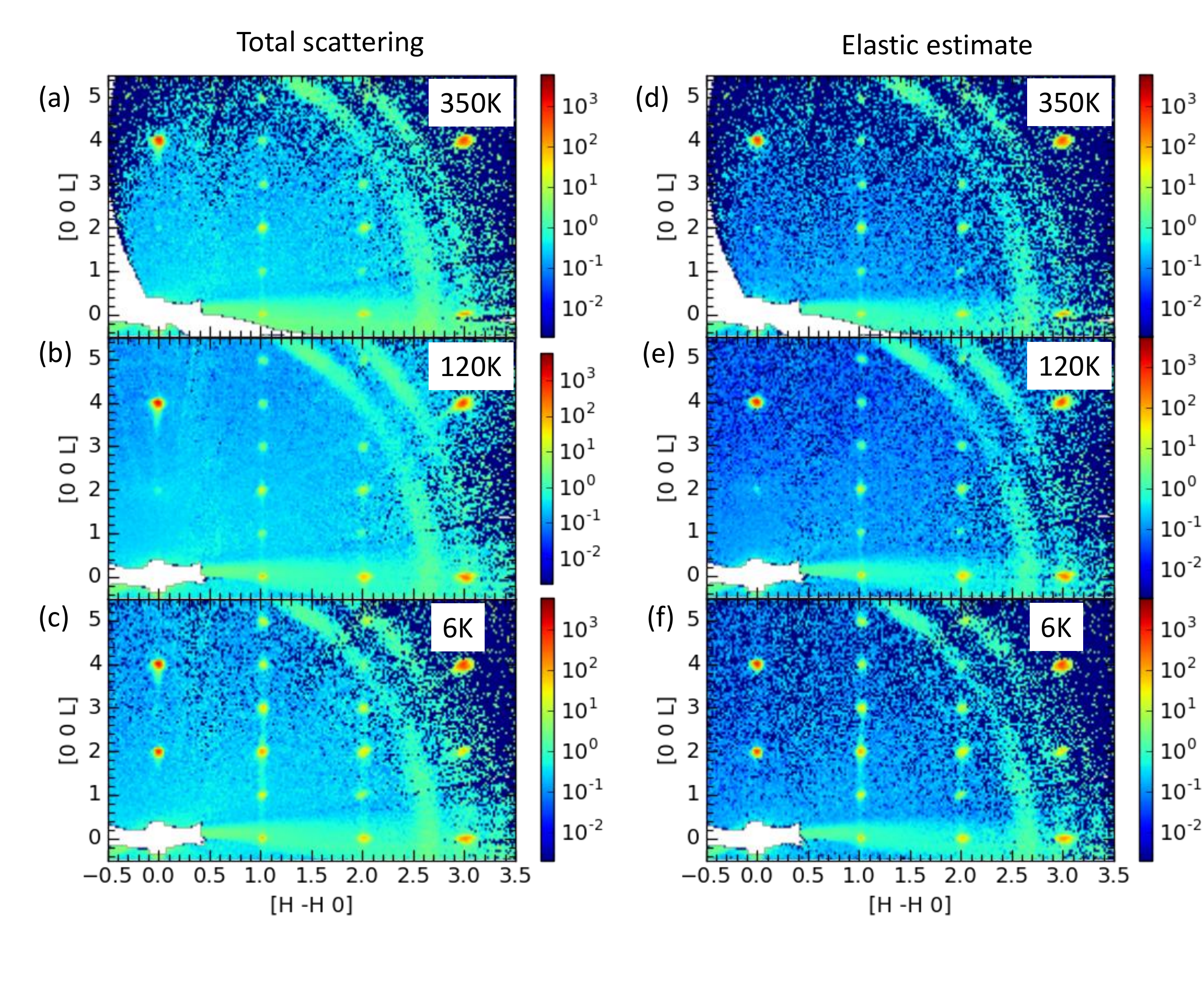}%
\caption{Neutron scattering intensity maps from the 2D detectors of CORELLI showing scattering in the H-HL plane (a,d) 6\,K, (b,e) 120\,K and (c,f) 350\,K. The two columns are the total scattering and the estimated elastic scattering. These 2D maps show the intensity from [-0.005 -0.005, 0] to [0.005, 0.005 0], i.e., a single bin, along the [H H 0] out-of-plane direction.}%
\label{Corelli1}%
\end{figure}

Diffuse magnetic scattering was investigated using the same crystal as the neutron diffraction measurements.  The crystal is approximately 7\,mg in mass, which is small when searching for diffuse scattering intensity.  As such, long counting times were employed.  In addition, the measurement is facilitated by the large moment on the Mn atoms.  The main text discussed diffuse scattering along L in the (H-HL) plane, which was strongest at 6\,K.  This diffuse scattering is centered around L=2, and is essentially absent for L $<$ 1 and L $>$ 3.  As discussed in the main text, the source for this scattering is not entirely clear. Given that this occurs for non-zero H, and is not present for large L, it potentially relates to a canting of the moments out of the $ab$ plane.  Additional data for this scattering plane, including the estimated elastic contribution, are shown in Fig.\,\ref{Corelli1}

The diffuse scattering at 6\,K is very anisotropic, occurring only along L and not along H.  Such one-dimensional diffuse scattering is sometimes observed in quasi-2D crystal structures due to the presence of stacking faults.  An important feature observed here is the lack of diffuse scattering for H = 0.  From a magnetic scattering perspective, only non-zero H reflections in this plane would possess scattering from a component of the moment along the $c$ axis.  Therefore, these results would be consistent with the presence of some canting along the $c$ axis that only has short coherence lengths.  This would be consistent with the average structure being that of a non-canted moment, and we recall that only a very small canted moment is observed when allowed during refinement of the single crystal neutron diffraction data at 4\,K.  Short coherence lengths of a canted moment could perhaps be caused by the competing exchange interactions.  The existence of a large J$_{1}$ promotes ferrimagnetic Mn1-Mn2-Mn1 units along the $c$ axis, and the coupling of these via competing interactions produces the long-range ferrimagnetic structure.  Thus, it may be possible that each of these units (or clusters of these units) has a certain canting angle (or lack thereof), but the coherence between them is somewhat weak.  It is also possible that a crystallographic defect on the Mn positions causes the diffuse scattering, which becomes observable through a strong magnetic contribution at low $T$.

\section*{Analysis of specific heat capacity data}

\begin{figure}[h!]%
\includegraphics[width=\columnwidth]{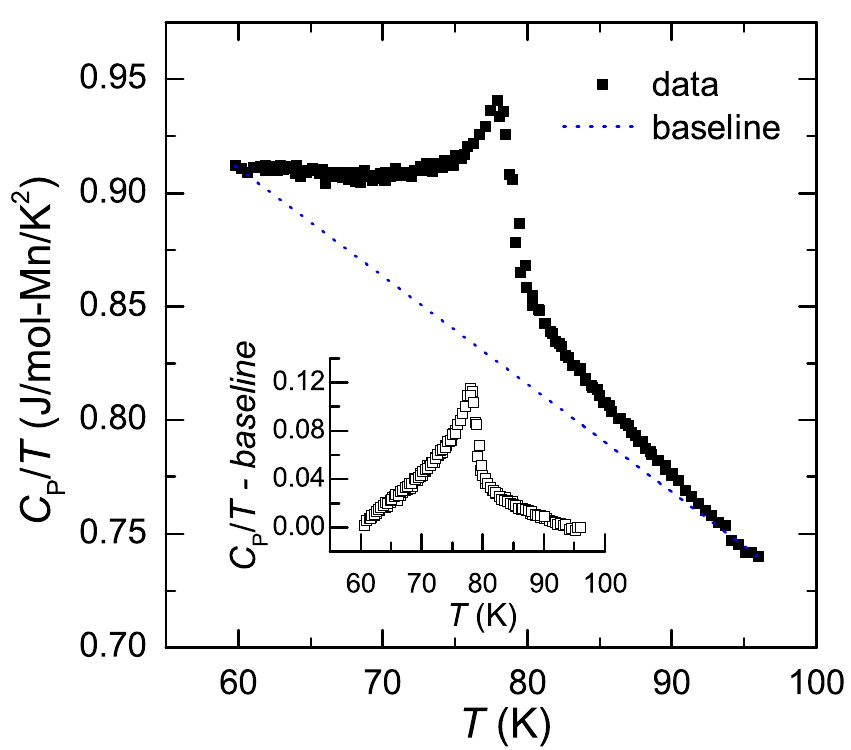}%
\caption{Specific heat data as $C_P/T$ versus $T$, the integration of which provides an entropy change.  To integrate over the anomaly near \TC without a phonon reference material, we have taken a straight line for the baseline and this will typically cause an overestimation of the entropy change across \TC.  The inset shows the data after removal of this baseline, which would essentially correspond to the magnetic contribution if an appropriate baseline were utilized}%
\label{CPFit}%
\end{figure}

As mentioned in the main text, we performed a fit of the specific heat capacity data to estimate the magnetic entropy change at \TC.  Importantly, this was done in a manner that would generally maximize any contribution (overestimation).  The reason for this is to show that even when overestimated, the entropy change across \TC is very small (about 8\% of anticipated, for the data and baseline shown).  The data utilized are shown in Figure \ref{CPFit}, and the inset shows the data after subtracting the baseline (same units as primary y-axis).

\begin{figure}[h!]%
\includegraphics[width=\columnwidth]{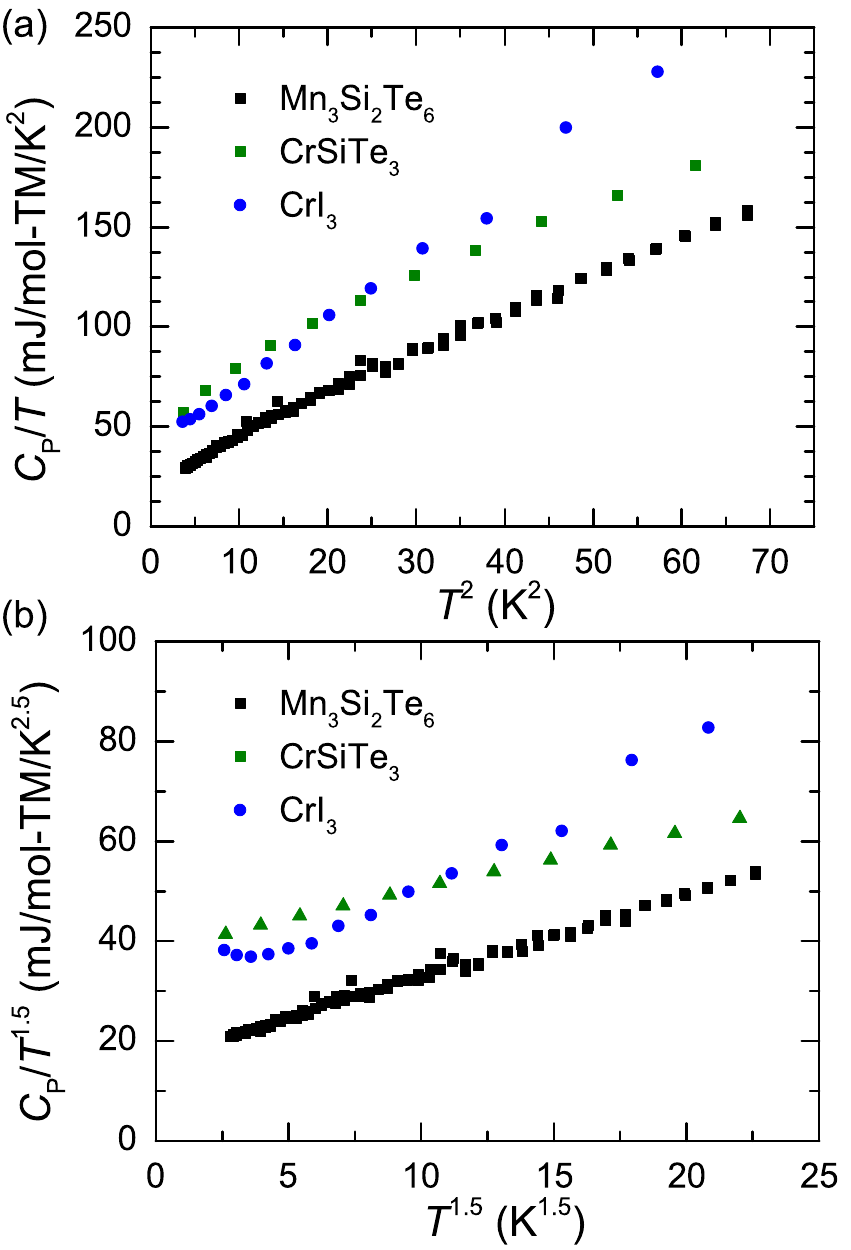}%
\caption{Low temperature specific heat capacity of \MST, CrSiTe$_3$ and CrI$_3$ are plotted assuming (a) $C_{\rm P}$ = $\gamma T$ + $\beta T^3$ formulation and (b) using a $C_{\rm P}$ = $\gamma\,' T^{1.5}$ + $\beta T^3$ representation. In the axis labels, the term TM = transition metal.  Here, $\beta$ is assumed to be related to the Debye temperature as usual, but $\gamma$ is not meant to represent the usual Sommerfeld coefficient because these materials are semiconducting and should not possess a significant electronic contribution to $C_{\rm P}$ at low $T$.  The data for CrSiTe$_3$ are taken from Ref.\,\citenum{Casto2015-SI} and data for CrI$_3$ are from Ref.\,\citenum{McGuire2015-SI}.}%
\label{CpMag}%
\end{figure}

To further examine the magnetic contribution to the specific heat, we have compared data for \MST with that of CrSiTe$_3$ and CrI$_3$ in Fig.\,\ref{CpMag}.  Both of these Cr-based compounds are ferromagnetic semiconductors and are quasi-2D with van der Waals gaps; CrSiTe$_3$ has \TC=33\,K and CrI$_3$ has \TC=61\,K.\cite{Dillon1965-SI,McGuire2015-SI}  While the magnetic order of these materials is different from that of \MST, they all possess relatively large magnetic contributions to the specific heat capacity at low $T$.  In Fig.\ref{CpMag}a, the data are plotted in the conventional $C_P/T$ versus $T^2$ manner.  This appears valid for CrI$_3$ and does an adequate job of describing the data for \MST, though the data for \MST have some curvature over the wide temperature range considered.   A more extreme curvature is found for CrSiTe$_3$, where non-linearity of $C_P/T$ versus $T^2$ is clearly evident.  Instead, CrSiTe$_3$ is better modeled using a magnetic contribution that is proportional to $T^{1.5}$ (Fig.\ref{CpMag}b).  This treatment also seems to work for \MST, but clearly fails for CrI$_3$.  At the lowest $T$, the data for \MST are well-described by a linear term, as shown in the main text.

\end{document}